\documentclass[portuguese,twoside,a4paper,12pt,openright]{book}
        \usepackage[applemac]{inputenc}
        \usepackage{amsthm} 
        \usepackage{amssymb}
        \usepackage{graphicx} 
        \usepackage{rotating} 
        \usepackage{multirow} 
        \usepackage{amsfonts}
        \usepackage{amsmath}
	\usepackage{float}
        \usepackage{colortbl}
        \usepackage{fancyhdr}
        \usepackage{textcomp}
        \usepackage{hyperref}
                \usepackage{picinpar}  
\newcommand{\HRule}{\rule{\linewidth}{0.5mm}}
\usepackage[avantgarde]{quotchap}

		\makeindex
		
\begin{document}
		\begin{titlepage}
		\center

\HRule
		\vspace*{0.5cm}
		{\Huge \bfseries \textsf{ASPECTS OF INFLATION\\ AND THE VERY EARLY UNIVERSE}\\}
		\vspace*{0,5cm}

\HRule  \vspace*{2,5cm}

		   \includegraphics[width=3cm]{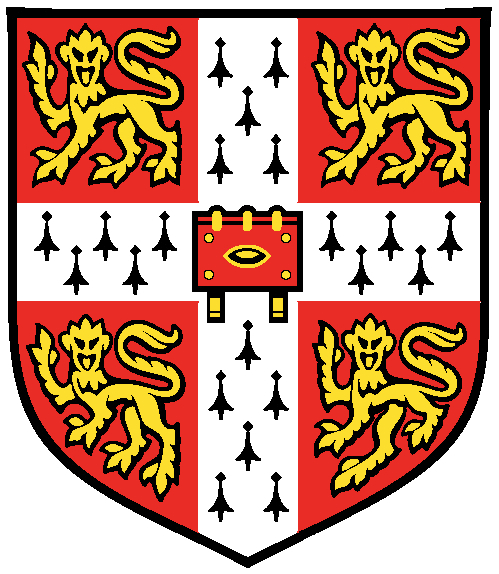}

		\vspace*{1cm}

		\begin{normalsize}
\begin{LARGE}
\textsf{I Raquel Henriques Ribeiro}
\end{LARGE}
\end{normalsize}
		\vspace*{2.5cm}

		\begin{footnotesize}
		This thesis is submitted to the University of Cambridge\\
		for the degree of Doctor of Philosophy \\

August 2012\\
\end{footnotesize}

		\vspace*{2cm}

		\begin{Large}
Girton College
\end{Large}\\
		Cambridge, UK
		\end{titlepage}

	\newpage
		\pagestyle{fancy}

	\thispagestyle{empty}
	\begin{center}
	\vspace*{\fill}
	\textsf{PhD Thesis \\ Supervisor: Anne-Christine Davis\\ Department of Applied Mathematics and Theoretical Physics - \textbf{DAMTP}\\
	University of Cambridge, UK}
	\end{center}
	
	\newpage
	\frontmatter
	\thispagestyle{empty}

	\vspace*{5cm}

	\begin{flushright}
	\textit{Aos meus avós, os meus maiores mentores. \\
	Áqueles que partiram, \\e demasiado cedo se tornaram poeira estelar, \\mas que em mim respiram e se perpetuam.}
	\vfill
\end{flushright}	

\renewcommand{\baselinestretch}{1.5}
\small\normalsize

\chapter{Abstract}


\chapter{Acknowledgements}

Writing up this thesis does not really represent, to me, the end, but rather an introspection moment. Coming to Cambridge has changed my life in so many ways that it is hard to summarise these in words. 

I would like to appreciate all the guidance and support of my supervisor, Prof. Anne Davis.

 Integrating backwards, I have always felt thrilled facing the big Cosmos surrounding us. Mamã, a ti te devo esta sede. My grandmother is the bravest, most intelligent person I know (and will probably ever know) who holds the ``straw" of life - to her I owe my stubbornness. 

Pedro, thanks for being here and there, then and now - your support was timeless and your presence was a safe candle in the way. At last, but certainly not the least, a word of gratitude to my beautiful, awesome friends: Linda, I will never forget our crazy escapes to the Royal Opera House - I will always remember those days with excitement; João, you were the best next-door officemate ever - thanks for our endless conversations and mid-afternoon breaks.

\tableofcontents
\listoffigures
\listoftables
	\renewcommand\chapterheadstartvskip{\vspace*{-5\baselineskip}}
\mainmatter  

\begin{savequote}[30pc]
Sometimes I think we're alone in the universe,\\ and sometimes I think we're not. \\In either case the idea is quite staggering.
\qauthor{Sir Arthur C. Clarke }
\end{savequote}

\chapter{Introduction}

\hrule
\vspace*{1cm}

The universe is an incredibly fascinating subject of study.
Ever since we gained the ability of logical thinking we have 
been staring at the sky bewildered by the enormous amount of space out there.
It is in the cosmos that we find the oldest and longest journey of all times: that of the 
primordial perturbation.

The first major discovery in cosmology over the last century 
was due to Edwin Hubble 
who observed that galaxies were moving apart from each other: the further
away they were the larger the speed of separation from us.
We learned the universe was not static, but expanding.
After this discovery much of the work in cosmology was 
theoretical.
In the last few decades, however, this has changed since a remarkable collection 
of high-precision satellites was assembled. 
In the last two decades, our knowledge of the universe 
has grown tremendously as
we have become acquainted with its extraordinary history.
There are essentially two important events that allowed this to happen.

First, the launch of the \textit{COBE} (COsmic Background Explorer) satellite\footnote{
				http://lambda.gsfc.nasa.gov/product/cobe/, April 2012.}
in 1989 established the 
beginning of measurements of the Cosmic Microwave Background 
Radiation \cite{Copeland:1993gs}, which we refer to as the CMBR from now on. 
The CMB was first discovered accidentally by Penzias 
\& Wilson in 1965, who collected the Nobel Prize in 1978.
This breakthrough triggered many questions about the CMBR properties.

\vspace*{0,5cm}

\begin{figwindow}[0,1,\includegraphics[width=8cm]
{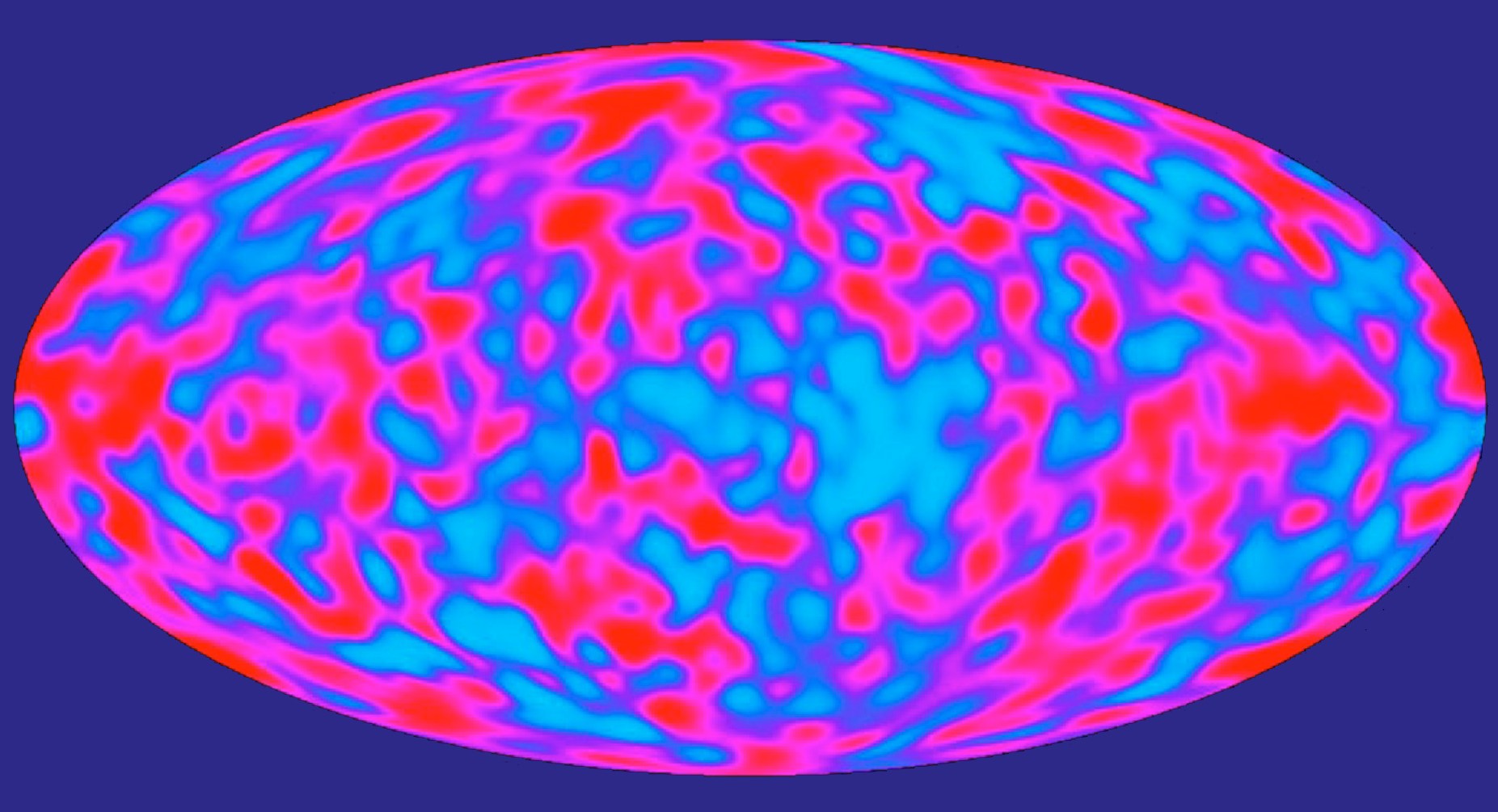}, {CMB 2-year data showing the temperature anisotropies in the microwave sky (image 
courtesy of the \textit{COBE} official website).}] 
\noindent  Measurements of the CMBR are purely statistical, and so is our knowledge of 
the universe in its infancy.
\textit{COBE} performed the very first measurements of the anisotropies of the CMBR, 
as small as 1 part in 1,000, in a background radiation of roughly
2.7K. 
With \textit{COBE} we learned that the CMBR has an almost perfect 
black-body spectrum.
\label{fig:cobe}
\end{figwindow} 
\vspace*{1,5cm}

It was with the follow-up mission of 
\textit{WMAP} (Wilkinson Microwave Anisotropy Probe)\footnote{http://map.gsfc.nasa.gov/, April 2012.}, launched in 2001, 
that the details of the statistics of the CMBR became known. 
By comparing figures 1.1 and 1.2
we appreciate the amazing improvement of precision 
in temperature measurements.
The details of the colour map in the second figure 
are undeniable evidence of the extraordinary advances in 
observational cosmology.

\begin{figwindow}[0,1,\includegraphics[width=8cm]
{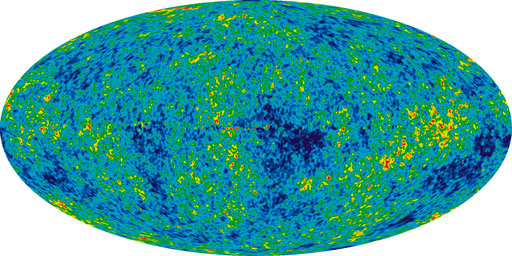}, {7-year \textit{WMAP} data of the microwave sky (image 
courtesy of NASA/WMAP science team).}] 
\noindent  These statistics are much more than just numbers. They 
encode detailed information about the early universe. 
They are remnants of an early, hot, violent epoch
imprinted in the microwave radiation---they reflect the microphysics of the early universe.
\label{fig:wmap}
\end{figwindow} 
\vspace*{1cm}

With the launch of \textit{WMAP} we entered the high-precision era of cosmology, 
and the story of fluctuations went from being purely 
theoretical to become observationally quantitative. This understanding builds up the  
\textit{early universe cosmology}.

Second, observations starting in 1998
from supernovae type IA showed that the universe today is
undergoing an accelerated expansion 
\cite{Perlmutter:1997zf, Riess:1998cb, Tonry:2003zg}. 
As standard candles, these supernovae are important 
tools to measure cosmological distances, which made this discovery possible.
This was a rather unexpected result. We would naturally expect that all the matter
in the universe would cause the expansion to slow down.
\begin{figure}
\begin{center}
\includegraphics[width=13cm]{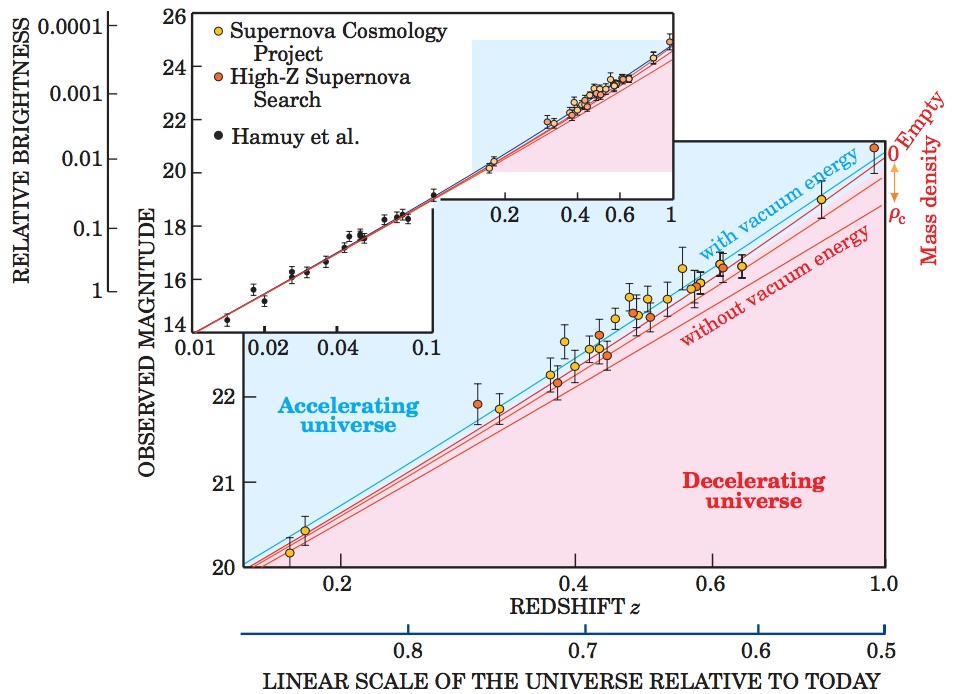}
\label{fig:supernovae}
\caption{Experimental data taken in 1998 for different values of redshift
 favours cosmological models where 
vacuum energy (dark energy) contributes to the total energy density in the universe.
The plot is taken from an article written by Saul Perlmutter which appeared in Physics Today
\cite{Perlmutter:physicstoday}.}
\end{center}
\end{figure}
To explain this strange behaviour we have put forward the existence 
of a mysterious force that, against gravity, 
drives cosmological structures apart, faster and faster. 
To understand this ``dark energy'' component in the Cosmos
is one of the main goals of the \textit{late universe cosmology}.

It is no accident that both these events have earned the recognition
of the Nobel Committee. 
Both branches of cosmology have brought together a 
global effort to understand our universe's history, 
from the early times to today.
In this thesis we will focus on the universe at early times.

The CMBR 
is the oldest memory of the 13.7 billion year-old universe
we live in. It is indeed the 
only fossilized record of microphysics in the early universe
which shows evidence of a young, vibrant universe.
The CMBR is the first light in the universe and its
fantastic journey started at the surface of last scattering, when 
the universe was about $380,000$ years old.
We measure the anisotropies in the CMBR as fluctuations in temperature, 
against an almost uniform temperature field. 
The statistics in the temperature field
can in turn be used to trace back what 
 mechanism generated them.
To understand the origin of these fluctuations in the temperature field, 
we need to discuss \textit{inflation}.

\section{FLRW cosmology}
Cosmology is based on the simple observation that the universe 
at large scales is roughly homogeneous and isotropic.
This is known as the \textit{Cosmological Principle}. 
To describe it we assume a
Friedmann--Lema\^{i}tre--Robertson--Walker space-time, given by the line element
\begin{equation}
ds^2= g_{\mu \nu} \d x^\mu \d x^\nu=-\d t^2 +a^2(t) \, \bigg( 
\dfrac{\d r^2}{1-k r^2}+ r^2 d\Omega_{\mathcal{S}^2}^2
\bigg)\ \ ,
\label{eq:flrw}
\end{equation}
where $t$ is the cosmological time, $a$ is the scale factor,  
$k$ measures the spatial curvature, and $d\Omega_{\mathcal{S}^2}^2$
denotes the line element of a 2--sphere.
The spatial sections of the 4-dimensional metric \eqref{eq:flrw} 
are homogeneous and isotropic. 
This asserts that wherever we look in the sky, at large scales, 
we observe more or less the same temperature 
field distributions. 
Different values of curvature $k$ correspond to different geometries:
$k=0$ corresponds to flat space, $k=-1$ corresponds to 
a 3--dimensional hyperboloid, and $k=1$ corresponds to a 3--sphere.
In Cartesian coordinates, for $k=0$, the line element above reduces to
\begin{equation}
ds^2=-\d t^2 +a^2(t) \, \d\vec{x}^2\ \ .
\label{eq:flrw2}
\end{equation}

\para{Conventions}In this thesis we use Greek letters $\{\mu, \nu, \cdots \}$
to denote space-time indices, whereas Roman letters
$\{a, b, \cdots \}$ are used for spatial indices. 
Equations are written in natural units, 
$c=1=\hslash$, and the reduced Planck 
mass, $M_P\equiv (8\pi G)^{-1/2}$, is set to unity.
The space-time metric is written as $g^{\mu\nu}$
and the signature is mostly plus $(-, +, +, +)$.
$\nabla$ denotes the space-time
	covariant derivative, and $\partial$ represents a spatial derivative.
	Dots denote differentiation
	with respect to cosmological time, 
	whilst primes denote 
	differentiation with 
	respect to conformal time (except when explicitly said otherwise).

It is usually assumed that the content of the universe
can be represented by a perfect fluid.
The energy-momentum tensor for a perfect fluid
with energy density $\rho$ and pressure density $p$ is given by
\begin{equation}
T^{\mu\nu}=(\rho+p)U^{\mu} U^{\nu} +p g^{\mu \nu} \ \ ,
\end{equation}
where $U^{\mu}$ is the 4-velocity of the fluid. If such fluid is
at rest in a geometry given by Eq. \eqref{eq:flrw}, and obeys the 
equation of state $p=w\rho$, then from the covariant conservation of the
energy-momentum tensor, we can find that
\begin{equation}
\dfrac{\d \rho}{\d t}+3H (1+w) \rho=0 \ \ ,
\label{eq:continuity}
\end{equation}
where $H \equiv \dot{a}/a$ is the Hubble parameter which
quantifies the rate of growth of the scale factor.

The solution to the continuity equation \eqref{eq:continuity} is
\begin{equation}
\rho \sim a^{-3(1+w)} \ \ .
\end{equation}
For pressure-free matter, as the universe expands
 the energy density is diluted $\rho\sim a^{-3}$, 
whereas for radiation $\rho\sim a^{-4}$. An interesting behaviour occurs for 
$w=-1$, which corresponds to a fluid with negative pressure. In this case, 
the energy density is constant as the universe changes size.
Nothing we know on Earth behaves this way.
We will come back to this odd feature shortly.

The dynamics of the scale factor
is controlled by Friedmann's equation
\begin{equation}
H^2 =\dfrac{\rho}{3}-\dfrac{k}{a^2} \ \ ,
\label{eq:friedmann}
\end{equation}
which depends on the total energy density and on the curvature of the spatial slices.
Eqs. \eqref{eq:continuity} and \eqref{eq:friedmann} are the 
fundamental equations for the dynamics of the scale factor, and 
imply 
Raychaudhuri's equation
\begin{equation}
\dfrac{\ddot{a}}{a} =-\dfrac{\rho}{6} \, (1+3w)\ \ .
\end{equation}
For a universe mostly made of baryonic matter it is clear that the expansion 
has a \textit{negative} acceleration. 
However, the present accelerated expansion
means that the energy density of the universe is dominated by 
atypical matter, so that $w<-1/3$. This determines the equation of 
state of a fluid which generates repulsive gravity.

\para{Energy density parameter}To consider several possible contributions to the energy density in the 
universe, it is common to define the density parameter
\begin{equation}
\Omega_i\equiv \dfrac{\rho_{i}}{\rho_C}\ \ ,
\end{equation}
where $\rho_C$ is the critical energy density for which 
the universe is flat. From Eq. \eqref{eq:friedmann}
we find that $\rho_C= 3H^2$. We can write the 
total density parameter as a sum over the contributions from 
the individual fluids which build 
up the total energy density in the universe:
\begin{equation}
\Omega \equiv \sum_i \Omega_i\ \ .
\end{equation}
In terms of this parameter, the Friedmann equation takes the form
\begin{equation}
H^2=H_0^2 \bigg(\sum_i \Omega_i a^{-3(1+w_i)}+\Omega_k a^{-2} \bigg)\ \ ,
\label{eq:friedmann1}
\end{equation}
where we are using the usual subscript `$0$' to denote
evaluation at the present time, when 
by convention $a(t_0)\equiv a_0=1$.
Above we defined $\Omega_k=-k/H_0^2$ to be 
the density parameter of the curvature.
Evaluating this equation today, we get
\begin{equation}
\sum_i \Omega_i +\Omega_k=1\ \ .
\end{equation}
This has a simple consequence: if the curvature of the universe
is observationally very small, that means that
\begin{equation}
\sum_i \Omega_i \equiv \Omega \approx  1 \ \ .
\end{equation}
Writing the equations in terms of $\Omega$ will prove 
to be particularly useful when discussing the flatness problem
of the Hot Big Bang paradigm.

\para{Conformal time}It is often convenient to define conformal time
\begin{equation}
 \tau = \int^{\infty} \dfrac{\d t}{a(t)}\ \ ,
\end{equation}
which runs from $-\infty$ to $0$. In these coordinates, 
the beginning of time $t=0$ corresponds to taking the limit when 
$\tau$ tends to $-\infty$. 
In terms of this time variable, 
the comoving distance travelled by a particle
corresponds to the distance measured as if 
we neglected the expansion of the universe.
Light rays will propagate in straight lines in the $(\tau, \vec{x})$ space
at 45º.
In this case, the geometry in flat FLRW space becomes much 
simpler
\begin{equation}
\d s^2= a^2(\tau) \big(-\d \tau^2 +\d \vec{x}^2 \big)\ \ .
\end{equation}
We see that the metric tensor $g^{\mu\nu}$ is 
conformally related to the Minkowski metric, $\eta^{\mu\nu}$,
via the scale factor, $a$, which is evolving in time. 

In conformal time, Friedmann's equation becomes
\begin{equation}
\mathcal{H}^2 \equiv a^2 H^2=a^2 \bigg(\dfrac{\rho}{3}-\dfrac{k}{a^2} \bigg)\ \ ,
\label{eq:friedmanntau}
\end{equation}
and Raychaudhuri's equation can be written as 
\begin{equation}
\mathcal{H}'=-\dfrac{\rho a^2}{3}\, (1+3w)\ \ ,
\label{eq:raytau}
\end{equation}
where the prime denotes derivative with respect to conformal time
and $\mathcal{H}$ is the Hubble parameter defined in conformal time.

\section{Big Bang Model}
Homogeneity and isotropy offer
a very simple description of 
the universe on large scales.
However, it brings conceptual problems 
to the Hot Big Bang paradigm when the universe is 
composed by conventional sources of matter, 
which we briefly describe next.

\para{The flatness problem}Cosmological observations show that the universe is approximately
flat today, with $k$ being almost zero. What are the implications of
this observation?
From Eq. \eqref{eq:friedmanntau} we find
\begin{equation}
\Omega -1 = k \mathcal{H}^{-2}\ \ .
\label{eq:omega_flat}
\end{equation}
We can show that the density parameter obeys the differential equation
\begin{equation}
\dfrac{\d \Omega}{\d t}=2H q (\Omega -1)\ \ ,
\end{equation}
where $q\equiv -\ddot{a}a/\dot{a}^2$ is the deceleration parameter. 
This equation governs the time evolution
of the density parameter, which for non-vanishing values of $H$ and $q$, 
indicates that $\Omega$ will generically be different from unity, 
unless $\Omega=1$ exactly always.
Also, differentiating Eq. \eqref{eq:omega_flat} with respect to conformal time
and using Eq. \eqref{eq:raytau}, we obtain
\begin{equation}
\dfrac{\d \ln \Omega}{\d \tau}= (1+3w) \mathcal{H} (\Omega -1)\ \ .
\end{equation}
Since $\Omega$ today is very close to $1$, writing $\Omega =1+\delta$, 
with $|\delta| \ll 1$, the last equation takes the form
\begin{equation}
\dfrac{\d \ln \delta}{\d \tau}= (1+3w) \mathcal{H}  \ .
\end{equation}
The solution to this equation is 
\begin{equation}
\delta (\tau) = \delta_0 \exp \bigg\{
(1+3w) \int_{\tau_0}^{\tau} {\mathcal{H} (\tilde{\tau})\d \tilde{\tau}}
\bigg\} \ \ .
\end{equation}
For fluids with $w>-1/3$, which in fact correspond to the baryonic matter 
and radiation we are familiar with, a universe with vanishing 
curvature presents a serious initial value problem---$k=0$ is not an attractor solution. With $\Omega$
diverging exponentially from unity, there is nothing that stops the universe from 
having a large curvature.

Another way of looking at this problem is to rewrite Eq. \eqref{eq:omega_flat}
as
\begin{equation}
\Omega^{-1}=1-\dfrac{3k}{\rho a^2}=1- \dfrac{3k}{\rho_0} \, a^{1+3w}(t) \ \ .
\label{eq:omegaflat}
\end{equation}
We observe that $\Omega$ indeed grows away from unity unless
$k$ is fine-tuned to be extremely close to $1$ initially
(under the assumption that the strong energy condition\footnote{The strong energy condition
													asserts that $\rho+3p>0$.}, 
$w>-1/3$, is obeyed).
A conservative approach based on observations of our universe
today, suggests that at the time of the Big Bang nucleosynthesis \cite{Lyth:2009zz}
\[|\Omega_{BBN}-1|\lesssim 10^{-16} \ \ . \]
Whilst it is possible that $\Omega$ started extraordinarily
close to $1$, it does require fine-tuning if the universe is
dominated by baryonic matter and radiation--this is the essence of the \textit{flatness problem}.

\para{The horizon problem}Wherever we look in the sky, the temperature is almost the same, 
up to very small differences. But why?
 Looking at two points 
which are separated by an angular distance larger than 2º, 
these points could not have been in 
causal contact and 
therefore they could have not shared any information. 
Yet, their temperature is almost the same.

To understand this problem better, it is useful to 
phrase the discussion in terms of the \textit{comoving particle horizon}. 
This is defined as the maximum comoving distance light could have travelled since the
universe was born, when $\tau \rightarrow -\infty$:
\begin{equation}
d_{ph}= \int_0^t \dfrac{\d\tilde{t}}{a(\tilde{t})}= 
\int_0^a \dfrac{\d \tilde{a}}{\tilde{a}^2 H}\ \ .
\label{eq:dph}
\end{equation}
It follows that events cannot be in causal contact when their distance is 
larger than the diameter of the comoving particle horizon.
For the discussion that follows, it is useful to write the 
comoving particle horizon in terms of the comoving Hubble radius, 
which evolves in time. 
We rewrite Eq. \eqref{eq:dph} as
\begin{equation}
d_{ph}= \int_0^a \d \ln \tilde{a}\  (\tilde{a}H)^{-1}\ \ .
\label{eq:dph-other}
\end{equation}
From Eq. \eqref{eq:friedmann1},
we observe that $d_{ph}$ increases monotonically in time
for $w>-1/3$:
\[ d_{ph} \sim a^{\frac{1}{2} (1+3w)}\ \ . \]
This has a puzzling consequence: scales which are entering the 
horizon now, were far outside the horizon when the universe 
became transparent---so, how can causally independent regions have the same
temperature? This is the \textit{horizon problem}.

\para{The relics problem}We know the early universe consisted of a hot, dense plasma, 
which was first dominated by radiation, and latter by baryonic matter.
Tracing back the history of the universe, we would expect
all the fundamental 
forces in Nature---electromagnetic, gravitational, 
weak and strong---to be unified.
As the temperature dropped, presumably a GUT phase transition would occur, 
which would break the symmetry amongst all the forces in Nature.
Such transition would have occurred at energies of order $10^{16}\mathrm{GeV}$, 
at which topological defects would have been produced.
These would have an energy density at present
far higher than that observed.

Since this expectation is 
in conflict with observations, monopoles and cosmic strings are usually
called unwanted relics.
We would therefore like to explain how these relics became so extraordinarily diluted.

\section{Inflationland}
\label{sec:inflationland}

Although there is still some debate about the origin of perturbations,
inflation is the most popular description of what we believe was
a dramatic event in the history of the universe. 
In the inflationary picture, when the universe 
was just a small fraction of a 
second old, it underwent a spectacular expansion phase, which pushed
cosmological scales far outside the horizon.
During this period space itself was expanding at a speed greater
than that of the light. 
This simple idea has important insights
when we try to explain the temperature distribution in the microwave sky.

So why is inflation such a good idea?
Inflation not only provides a mechanism by which
the problems of the standard cosmological model 
are ameliorated. When combined with Quantum Mechanics, 
it also explains the origin of the temperature perturbations
in the sky and the seeds of large scale structure.
It relates the physics of the very small and the very large.
It does this by assuming a very simple hypothesis: 
if the expansion rate of the universe was accelerated, 
\begin{equation}
 \ddot{a}>0 \ \ ,
\end{equation}
for at least 60 e-folds, then the comoving Hubble radius 
decreased in time:
\begin{equation}
\dfrac{\d }{\d t} (aH)^{-1}<0\ \ .
\end{equation}
Therefore
the observable universe went from being 
inside to outside the horizon
by the time inflation ends.

To understand how inflation worked, we focus our attention 
in Eq. \eqref{eq:dph}. If, during inflation, 
the Hubble parameter $H$ is roughly constant, 
then the comoving Hubble radius $(aH)^{-1}$ decreases in time. 
The largest contribution to the integrand in Eq. \eqref{eq:dph}
arises at early times, when scales were in causal contact, 
although they do not appear in causal contact now.

We see that the accelerated expansion of the universe requires the 
strong energy condition to be broken, and that at those 
early times the energy density in the universe
was dominated by a fluid with equation of state
\[ w \equiv \dfrac{p}{\rho}<-1/3\ \ . \]
The original formulation of inflation was put forth by Alan Guth in 1980
\cite{Guth:1980zm}
and marks an exploration into `inflationland.'
Let us see how it can ameliorate the problems
in the cosmological model.

First, the flatness problem. From Eq. \eqref{eq:omegaflat} it 
follows that if $w<-1/3$, then the evolution of $\Omega$ 
dictates it should reach the attractor solution $\Omega \rightarrow 1$.
Even if the curvature in the early universe was 
large, it would be greatly diluted away by the accelerated expansion.

Second, the horizon problem. This puzzle is resolved 
when we use the fact that the comoving Hubble radius
decreased during inflation. 
This means that scales that re-enter the horizon after 
inflation ended, were \textit{once} inside the horizon---they were
therefore once in causal contact. With this assumption, 
there is no surprise in the universe being composed of 
incredibly homogeneous patches even at 
large angular separations.

Finally, with inflation relics are naturally diluted away
since the accelerated expansion induces a dramatic 
decrease in the energy density of relic particles and topological defects.

With this positive outcome of assuming that a period of inflation
took place in the early universe, let us explore this epoch more deeply.
If the Hubble parameter $H$ varies very slowly in time, then 
a period of quasi-de Sitter inflation is possible. In the limit when $H$ is strictly
a constant, it follows that
\begin{equation}
a \sim e^{H t}\ \ .
\end{equation}
This can be described by a fluid with equation of state $w=-1$, 
and the scalar field associated with it is called the \textit{inflaton}. 
The inflaton is a field whose 
energy density dominates the total energy density in the early universe, 
and generates an exponential expansion of the scale factor. 
To describe such period one usually introduces a slow-roll parameter, defined by
\begin{equation}
\varepsilon \equiv -\dfrac{\dot{H}}{H^2}\ \ ,
\label{eq:defepsilon}
\end{equation}
which must obey $\varepsilon \ll 1$.
For inflation to be efficient, one
generically requires at least $60$ e-folds of expansion, 
where the e-folding number
\begin{equation}
N\equiv \int_{a_i}^{a_f}\d \ln a = \int_{t_i}^{t_f} H(t) \d t
\label{eq:efolds}
\end{equation}
is a measure of the expansion of the size of the universe, 
from $a_i$ to $a_f$. Its usefulness as a unit of time 
was first noticed in a paper by Sasaki 
\& Tanaka \cite{Sasaki:1998ug}.
To ensure that $\varepsilon$ stays very small 
during inflation, one introduces a further slow-roll parameter
\begin{equation}
\eta \equiv \dfrac{\dot{\varepsilon}}{\varepsilon H}\ \ ,
\label{eq:defeta}
\end{equation}
which is required to obey $|\eta|\ll 1$.

\subsection{A simple inflation model}
Which fluid with negative pressure could have sourced inflation?
It certainly could not have been radiation or baryonic matter, 
for which the pressure is positive definite.
To answer this question, let us take the simplest inflation model, 
involving only one scalar field, the inflaton, which we assume 
is homogeneous. This model is described by 
\begin{equation}
S= \int \d^4 x \sqrt{-g} \bigg\{\dfrac{1}{2}R -\dfrac{1}{2} (\partial \phi)^2
-V(\phi)\bigg\}\ \ .
\label{eq:actionsimple}
\end{equation}
In this action $\phi$ is only minimally coupled to the 
curvature of space-time.
The scalar field has a canonical kinetic term and is subject 
to a self-interacting potential, $V(\phi)$. Different 
models have different potentials. For now, we will assume an arbitrary $V(\phi)$.

A typical example is depicted in the following schematics.
\begin{figwindow}[0,1,\includegraphics[width=10cm]
{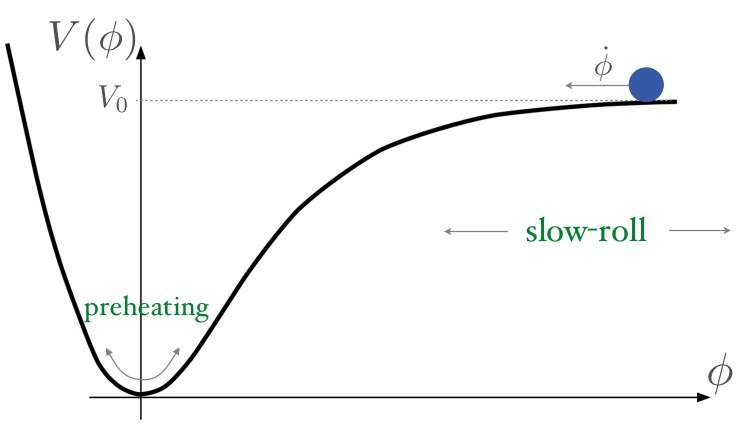}, {Simple (small-field) inflation potential
and the dynamics of inflation.}]
\noindent  During the slow-roll phase, the universe is inflating, 
and the potential is roughly constant, $V_0$.
This will be true until the potential energy of the scalar field 
$\phi$ no longer dominates
the energy density in the universe, and is unable to source an accelerated 
expansion. \label{fig:simplepotential}
\end{figwindow} 
\vspace*{1cm}
\noindent When this happens, the inflaton field speeds up and approaches the 
minimum of the potential. It then starts oscillating about the minimum, 
producing matter particles in the process, if it is coupled to other degrees of 
freedom. This process is called \textit{reheating}, and we will study it in detail 
in chapter \ref{chapter:six}.
To understand the slow-roll approximation better, 
we write the equations of motion for the scalar field.
Assuming an FLRW background \eqref{eq:flrw}, these are
\begin{equation}
\ddot{\phi}+3H \dot{\phi} +\dfrac{\partial V}{\partial \phi}=0\ \ .
\label{eq:simpleeom}
\end{equation}
The energy-momentum tensor can be found by varying the action 
\eqref{eq:actionsimple} with respect to the metric tensor, giving
\begin{equation}
T_{\mu \nu}= \partial_\mu \phi \partial _\nu \phi -g_{\mu\nu} \bigg\{ 
\dfrac{1}{2} (\partial \phi)^2-V(\phi)
\bigg\} \ .
\end{equation}
Interpreting the components of the energy-momentum tensor
as those of a fluid, we find that the energy density and pressure are 
given by
\begin{subequations}
		\begin{align}
\rho &= \dfrac{1}{2}\dot{\phi}^2 +V(\phi)\\
p &= \dfrac{1}{2}\dot{\phi}^2 -V(\phi)\ \ .
\label{eq:pressure-chapter1}
		\end{align}
\end{subequations}
As a result, the fluid has equation of state $w=-1$ if one neglects the kinetic energy 
compared to the potential energy. 
Moreover, if the potential is approximately constant
then the energy density in Friedmann's equation \eqref{eq:friedmann} 
is roughly constant, sourcing a de Sitter 
universe, and 
\begin{equation}
3H^2 \simeq V_0\ \ .
\end{equation}
This can be interpreted upon inspection 
of the action \eqref{eq:actionsimple}.
If $V(\phi)\simeq V_0$ the energy density of the fluid acts 
as a cosmological constant
and effectively sources inflationary expansion.
Guth referred to this as the 
``ultimate free lunch'' since despite the expansion of the universe, 
the energy density of the fluid was kept constant, as if it were sourcing it 
out of nothing.

The slow-roll approximation asserts that 
we can neglect the second time derivative of the field
in Eq. \eqref{eq:simpleeom} compared to the first time-derivative.
Then 
\begin{equation}
\dot{\phi}  \approx -\dfrac{1}{3H} \dfrac{\partial  V}{\partial \phi}\ \ .
\end{equation}
In terms of the features of the potential, we can also define 
the slow-roll parameters as follows:
\begin{subequations}
		\begin{align}
\varepsilon_V &= \dfrac{1}{2} \bigg( \dfrac{\partial V/\partial\phi}{V}   \bigg)^2 \ \ 
\textrm{and}\\
\eta_V &=  \dfrac{\partial^2 V}{\partial \phi^2}\ \ .
		\end{align}
\end{subequations}
These coincide with the definitions \eqref{eq:defepsilon} and \eqref{eq:defeta} 
up to slow-roll corrections.
We observe that the requirements $\varepsilon, |\eta| \ll 1$ translate 
into a potential that is almost flat for a large range of $\phi$, 
making its curvature very small. This is the slow-roll period 
illustrated in figure 1.4.

Slow-roll inflation lasts until $\varepsilon\sim 1$ and the kinetic energy
of the scalar field becomes comparable to the 
potential energy---the potential becomes steeper and the scalar field speeds up, 
as described above.
It is during the slow-roll phase that the scalar field fluctuations
become imprinted in the CMBR---we will discuss this in detail in \S\ref{subsec:seeding-pert}.

\subsection{Beyond the simplest inflation models}
\label{subsec:beyond-simple-inflation}
Inflation offers a simple, dynamical solution to the cosmological problems, 
but the exact mechanism driving the process of inflationary 
expansion is still rather speculative. This is because the precise physics of inflation
is essentially unknown and it is a matter of dispute whether we will ever 
be able to fully understand it.

Traditionally, we start with the simplest 
scenarios, like the action \eqref{eq:actionsimple}. We then add 
more ingredients to the action and try to understand their distinctive 
signatures in the CMBR in the hope to be able to distinguish between them
once appropriate observational data is available. 
There are a number of ways of generalising the action \eqref{eq:actionsimple}.
In what follows we mention some of the possible extensions.
\begin{enumerate}[i.]
\item \textit{inflation with modified gravity}. 
This can be achieved by assuming
a non-minimal coupling to the curvature scalar. 
These are the so called scalar-tensor theories, described by
\begin{equation}
S=\int d^4 x \sqrt{-g} \bigg\{
\dfrac{f(\phi)}{2} \ R -\dfrac{1}{2}(\partial \phi)^2 -V(\phi)-\mathcal{L}_{mat}
 \bigg\}\ \ ,
 \label{eq:action-scalar-tensor-theories}
\end{equation}
where $\mathcal{L}_{mat}$ defines the Lagrangian density of the matter sector.
This action can be rewritten in the form of Einstein's gravity by performing 
a conformal transformation
\begin{equation*}
g_{\mu\nu} \rightarrow \Omega^2 (t,\vec{x}) \ g_{\mu\nu}
\end{equation*}
where the conformal factor is appropriately chosen \cite{Hirai:1993cn}
\begin{equation*}
\Omega^2 = f(\phi)\ \ .
\end{equation*}
After applying this transformation to the action \eqref{eq:action-scalar-tensor-theories},
we say the theory of gravity is written in the Einstein frame.
With this transformation, the space-time curvature is modified and the 
Ricci scalar becomes \cite{Wald:1984rg}
\begin{equation*}
R \rightarrow \dfrac{1}{\Omega^2} \bigg\{R+6\ \dfrac{\Box \Omega}{\Omega} \bigg\}\ \ ,
\end{equation*}
with $\Box$ being the d'Alembertian, 
$\Box\equiv \frac{1}{\sqrt{-g}} \partial_\mu \big( \sqrt{-g} g^{\mu\nu} \partial_\nu \big)$.
To write the action in a more canonical way, we can further apply field redefinitions
\begin{equation*}
\d \phi \rightarrow \sqrt{\dfrac{2f+3f'^2}{4f^2}} \ \d \phi\ \ \
\textrm{and} \ \ \ V\rightarrow \dfrac{V}{f^2}\ \ ,
\end{equation*}
where $f'\equiv \partial f/\partial \phi$. Given these transformations, 
the action becomes
\begin{equation}
S=\int d^4 x \sqrt{-g} \bigg\{
\dfrac{1}{2} \ R -\dfrac{1}{2}(\partial \phi)^2 -V(\phi)-\mathcal{\tilde{L}}_{mat}
 \bigg\}\ \ ,
\end{equation}
with $\mathcal{\tilde{L}}_{mat}$ denoting the modified matter Lagrangian density.
Therefore, the modification of gravity observed in the action through the function $f(\phi)$
appears in the Einstein frame as a modification of the original
Lagrangian of the matter sector.
Nevertheless, whatever frame is chosen, the physics should stay the same.
These theories have gained interest in resent years in the light of chameleon field
theories (see, for example, Ref. \cite{Brax:2008hh}).

Other possibilities which fall under this class exist: Jordan--Brans--Dicke 
theory of gravity \cite{Brans:1961sx} 
(which also includes non-canonically normalized scalar fields), 
$f(R)$ gravity \cite{Nojiri:2006ri} 
(which can include higher powers of the Ricci scalar), 
and modified gravity in braneworld scenarios 
\cite{Brax:2004xh}.
For a recent review on modified gravity theories and its cosmological 
implications see Ref. \cite{Clifton:2011jh}.

\item \textit{inflation with non-canonical kinetic terms}.
This class of theories also goes by the name of $k$-inflation
\cite{ArmendarizPicon:1999rj}.
 In contrast with 
the action
\eqref{eq:actionsimple} it is not the potential energy of $\phi$ 
that supports a de Sitter
phase, but rather the non-canonical structure of the kinetic term. 
The action takes the form
\begin{equation}
S=\int \d^4 x \sqrt{-g}\bigg\{
\dfrac{1}{2}R +P(X,\phi)
\bigg\}\ \ ,
\label{eq:pxphiaccao}
\end{equation}
and it was initially proposed by Armend\'{a}riz-Pic\'{o}n, Damour \& Mukhanov.
Here $P(X,\phi)$ is an arbitrary function of the field profile $\phi$ and its
first derivatives through $X\equiv -(\partial \phi)^2$. We recover 
action \eqref{eq:actionsimple} by choosing
$P(X,\phi)=-1/2(\partial \phi)^2-V(\phi)$;
more generically this action reproduces different models for different 
choices of the function $P(X,\phi)$.
Without loss of generality, we can write
\begin{equation}
P(X,\phi)=\sum_n c_n (\phi) \ \dfrac{X^n}{\Lambda^{4n-4}}\ \ ,
\label{eq:PXphigeneral}
\end{equation}
where $\Lambda$ denotes some high-energy scale.
Slow-roll inflation is obtained when $\Lambda^4\gg X$. 


This action will play an important r\^{o}le in this thesis and we will 
explore its phenomenology in detail in the subsequent chapters. 
Here, we shall only mention that it can
support inflation even in the presence of very steep potentials.
The most striking general feature of the non-canonical models, however,
is that the inflationary dynamics and the consequences for 
observables becomes quite non-trivial. 
This corresponds to the limit $\Lambda^4 \sim X$ in 
Eq. \eqref{eq:PXphigeneral}. This limit raises one concern:
radiative corrections can induce large renormalizations of 
the coefficients $c_n(\phi)$, which can lead to a ill-defined 
quantum field theory. In this case, only if the theory is equipped 
with a protective symmetry can we escape the problem of radiative 
instability. Of all the higher derivative single-field models, 
Dirac--Born--Infeld \cite{Silverstein:2003hf} 
and galileon inflation \cite{Burrage:2010cu} are the only 
established examples where non-renormalization theorems apply.
We will be more precise about this statement in the subsequent chapters.

Whether these are the only radiatively stable models
is a very different query---we have nothing to say about this in this thesis.
We are rather interested in understanding what are the consequences
for observables 
of taking a generalised $P(X,\phi)$ action.

\item \textit{multi-field inflation.} In this case, there are multiple sectors in the action
in addition to the inflaton field.
This is a generic feature of the theory when
embedding inflation in a more 
fundamental parent framework, like string theory. Indeed, effective field theories
(examples of work in this area are 
Refs. \cite{Cespedes:2012hu, Achucarro:2012sm, Avgoustidis:2012yc}) and
supergravity theories (see, for example, Refs. \cite{Binetruy:2004hh, Choi:2004sx}) 
usually contain multiple degrees of freedom.
If all the fields play a r\^{o}le in the 
inflationary dynamics, then a single-field effective picture is not possible 
and the dynamics becomes more intricate then any single-field inflation model.
Early studies of models involving more than one degree of freedom 
include works by Linde \cite{Linde:1984ti}, Kofman \& Linde \cite{Kofman:1986wm},
and Silk \& Turner \cite{Silk:1986vc}.
\end{enumerate}

In this thesis we will focus on the perturbation theory of 
generalised single-field models, starting from the action
\eqref{eq:pxphiaccao}. We will not attempt to investigate 
multi-field inflation models or
modified gravity models in the sense described above, although
galileon models arise naturally in the decoupling 
limit of massive gravities 
\cite{deRham:2010ik, deRham:2010kj, Hassan:2011hr}---we will
discuss these in chapters \ref{chapter:three}
and \ref{chapter:four}.

\subsection{Seeding perturbations}
\label{subsec:seeding-pert}
We now turn to the question of how inflation can microscopically 
explain the origin of Large Scale Structure (LSS).
For this we need to consider the quantum-mechanical 
behaviour of the `single-clock' in the universe during inflation.
Such r\^{o}le is played by the inflaton field which 
sets the initial conditions for the evolution 
of perturbations, as originally explained by 
Guth \& Pi \cite{Guth:1982ec}, Hawking \cite{Hawking:1982cz},
and Bardeen, Steinhardt \& Turner \cite{Bardeen:1983qw}.

Let us first revisit figure 1.4 of
a typical schematics of a single-field 
inflationary model.
\begin{figwindow}[0,1,\includegraphics[width=9cm]
{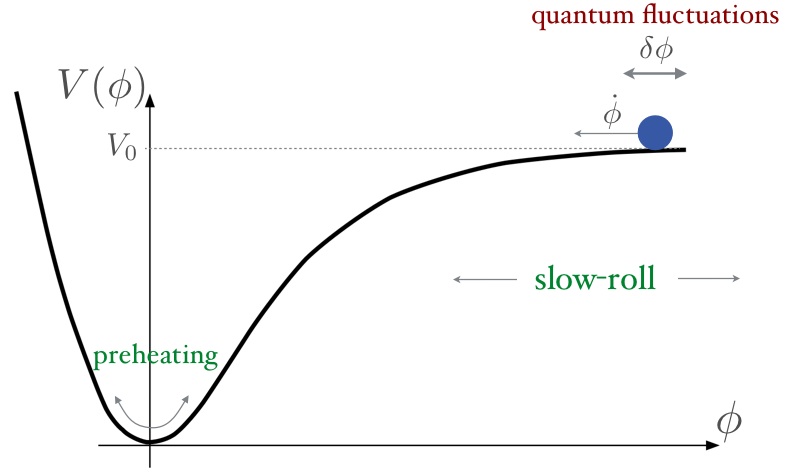}, {As inflation proceeds, the inflaton 
undergoes quantum fluctuations which are the seeds for density perturbations.}] 
\noindent  Going back to the idea that during inflation 
the comoving Hubble radius decreases, then quantum 
fluctuations with origin on sub-horizon scales will exit the horizon.
While on super-horizon scales, such fluctuations freeze until they are able 
to re-enter the horizon at late times. 
\label{fig:potentialsimplequantum}
\end{figwindow} 
\vspace*{1,7cm}

So how do the quantum fluctuations
translate into fluctuations in the density field, which can then respond 
to the gravitational pull and form bound structures?
Once created, the small, inhomogeneous inflaton perturbations, 
$\delta \phi (t,\vec{x})$,
will induce slightly 
different arrival times of the expectation value $\phi$ at the bottom of the potential $V(\phi)$.
As can be seen in the figure, 
regions in space where $\delta \phi$ is positive will remain 
potentially dominated for longer, whereas regions where $\delta \phi$ is 
negative will stop inflating sooner. Since $\phi$ is the only available 
clock in the universe (in single-field models), 
we can say that fluctuations $\delta \phi$ control 
the time differences at which inflation ends
through
\begin{equation}
\delta t = \dfrac{\delta \phi}{\dot{\phi}}\sim \dfrac{H}{\dot{\phi}}\ \ ,
\end{equation}
where hypersurfaces of constant time are also hypersurfaces of 
constant energy density, and therefore constant $H$.
After inflation ends and reheating proceeds, $H\sim 1/t$, and 
fluctuations in the expansion rate of the universe are mapped into
density inhomogeneities, as follows
\begin{equation}
\dfrac{\delta t}{t} \sim \dfrac{\delta H}{H} \sim \dfrac{\delta \rho}{\rho} \sim 
H \ \dfrac{H}{\dot{\phi}}\ \ .
\end{equation}

This way, the inflaton quantum perturbations
seeded structure during the early universe, 
which have later grown to form galaxies and clusters.
What is most impressive still is that the small 
inhomogeneities produced during inflation 
were imprinted in the early plasma, which cooled down and
became the CMBR after photon decoupling.
Today we observe these imprints in the microwave sky as temperature anisotropies, 
which produce the beautiful map depicted in figure 1.2.

\section{$\Lambda$CDM: the concordance model}

Measuring observables faces a tremendous challenge:
degeneracy. 
Given a set of data, different observables can compete 
and their effect might be so tied up in our measurements, that it 
is hard to find constraints for each individual observable.
To attempt to solve for this pressing problem, 
cosmologists need to have access to various, often complementary, 
sets of data.

The BOOMERanG (Balloon Observations Of 
Millimetric Extragalactic Radiation and Geophysics)
experiment\footnote{http://cmb.phys.cwru.edu/boomerang/, April 2012.} 
measured in 1997 the CMBR anisotropies, 
determining the value of Hubble's constant, $H$, and the fractional density of the 
universe, $\Omega$. 
It has also provided evidence in agreement with supernovae 
data on the present accelerated expansion, 
as announced in Refs. \cite{Melchiorri:1999br, deBernardis:2000gy}.
Other experiments include SDSS (Sloan Digital Sky Survey)\footnote{http://www.sdss.org/, April 2012.}, 
which is a ground-based telescope, that targets quasars and galaxies distributions, 
and places constraints on the cosmological parameters.

The best fit for the content of the universe today given
by \textit{WMAP} and BAO\footnote{BAO stands for Baryonic Acoustic Oscillations.} 
data \cite{Komatsu:2010fb, Larson:2010gs} 
is in table \ref{table:lambdacdm}.
The concordance model results from a significant component of the 
energy density in the universe being a 
cosmological constant, hence the name $\Lambda$CDM.

Observations indicate that the universe is made up of four basic components:
radiation, dark matter, baryonic matter, and cosmological constant (or dark energy).
The largest component is dark energy with a proportion 
of about 73\% of all there is.
Most of the universe is made up of something we cannot see and 
do not understand---we are only familiar with 4\% of the 
entire universe's budget. 
All the matter, $m$, in the 
universe is either baryonic matter, $b$, or cold dark matter $c$:
\[ \Omega_m=\Omega_b+\Omega_c \ \ . \]

Including supernovae data, cosmologists have learnt that the 
dark energy has equation of state
\[ w_{\Lambda}=-0.980 \pm 0.053 \ \ . \]
The data undeniably points to a flat universe.

\begin{table}[htpb]

	\heavyrulewidth=.08em
	\lightrulewidth=.05em
	\cmidrulewidth=.03em
	\belowrulesep=.65ex
	\belowbottomsep=0pt
	\aboverulesep=.4ex
	\abovetopsep=0pt
	\cmidrulesep=\doublerulesep
	\cmidrulekern=.5em
	\defaultaddspace=.5em
	\renewcommand{\arraystretch}{1.6}
	\begin{center}
		\small
		\begin{tabular}{Qq}

			\toprule
			\textrm{parameter}
			&
			\multicolumn{1}{c}{ best fit (mean)}
			\\
			\cmidrule(l){1-2}

			\Omega_b & 
			0.0458 \pm 0.0016 
			\\[2mm]

			\cmidrule{1-2}

			\Omega_c  & 
			0.229 \pm 0.015 
			\\[2mm]

			\cmidrule{1-2}

			\Omega_\Lambda h^2 &  
			0.725 \pm 0.016
			\\[2mm]

			\cmidrule{1-2}

			\Omega_k  &  
			-0.0133 - 0.0084 
			\\
			[2mm]
			
			\cmidrule{1-2}
			
			\Omega_m h^2 & 
			0.1352 \pm 0.0036 
			\\[2mm]
			
			\cmidrule{1-2}
			
			H_0 & 
			070.2 \pm 1.4 km/s/Mpc 
			\\[2mm]
			
			\cmidrule{1-2}
			
			t_0 & 
			13.76 \pm 0.11 Gyr 
			\\ 
 			\bottomrule
	
		\end{tabular}
	\end{center}
	\caption{The content of the universe given by direct measurements 
	and induced parameter values obtained from \textit{WMAP} \& BAO data
	\cite{Komatsu:2010fb, Larson:2010gs}. $t_0$ denotes the age of the universe.
	\label{table:lambdacdm}}
	\end{table}

We have intentionally removed all the data related to the CMB anisotropies
since we will discuss it in the next chapters.

\section{\textit{Planck}: the ultimate `thermometer'}
As described before, the last few decades have seen remarkable
advances in cosmology. These were only made possible owing to 
an unparalleled progress in observational cosmology. 
In May 2009 the \textit{Planck} 
satellite\footnote{http://www.esa.int/SPECIALS/Planck/index.html, April 2012.} 
was launched in an attempt to exhaust the amount of information one could extract from the CMB.

\vspace*{0.5cm}

\begin{figwindow}[0,1,\includegraphics[width=8cm]
{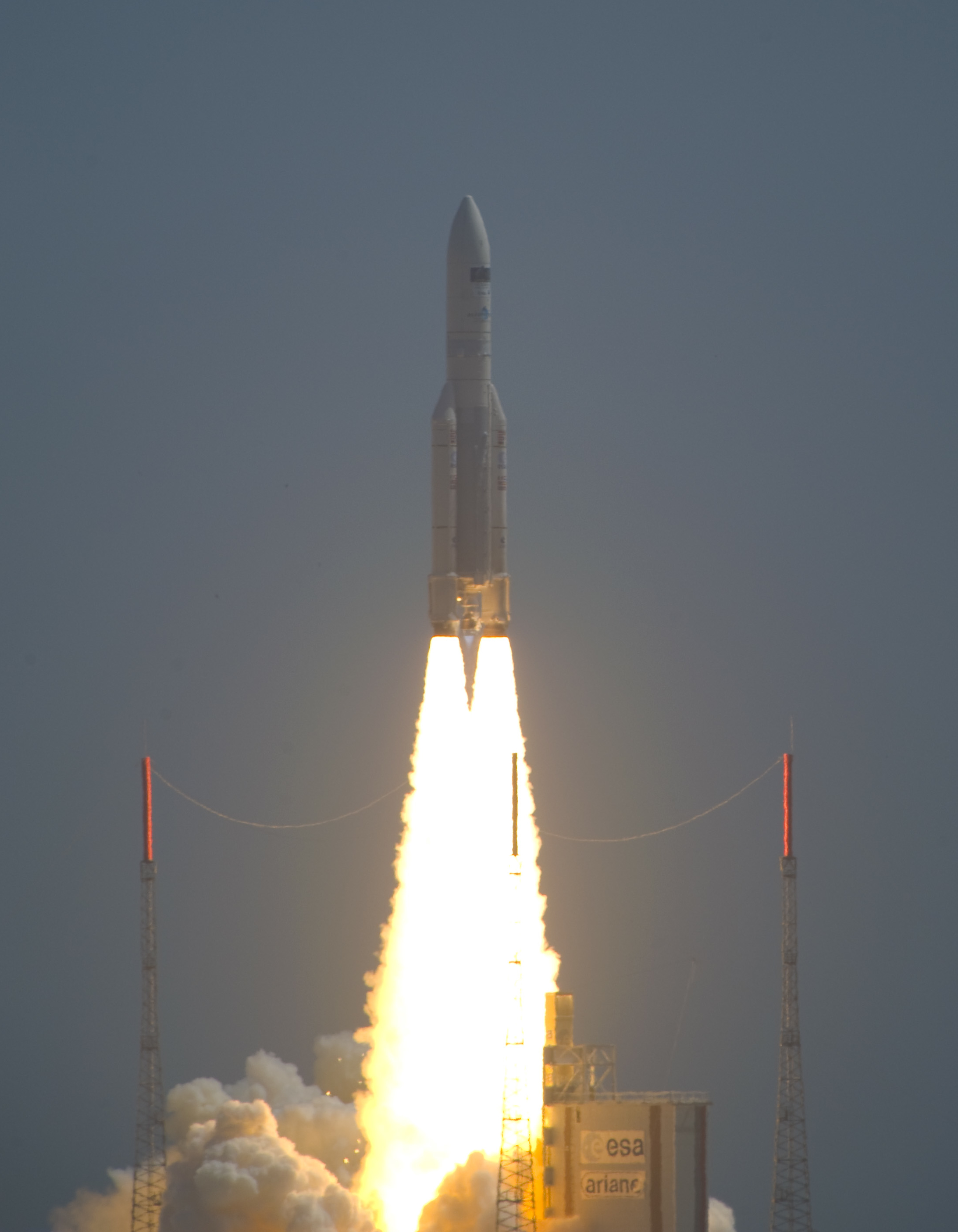}, {The launch of the \textit{Planck} satellite and the Herschel telescope
by Ariane 5
(image courtesy of ESA-CNES-Arianespace, Optique Vidéo du CSG, P. Baudon).}]
\noindent One of the main objectives of this mission was to 
enable a rigorous test of the early universe models against observation.
The sensitivity of the temperature measurements that \textit{Planck} offers 
is simply staggering: one part in one million. 
Having a direct focus on the detection of
primordial gravitational waves and specification of its spectrum, 
\textit{Planck} will also allow the most detailed study 
of the scalar spectrum of perturbations yet performed. 
In particular, with \textit{Planck} we expect to obtain very 
precise data on the higher-point correlations of the temperature field,
which we can map backwards to draw the features of the 
primordial curvature perturbation.
\label{fig:wmap1}
\end{figwindow}
 
\vspace*{1cm}

It is anticipated that 
the major problem in analysing data will be to efficiently
remove the foreground. It is therefore quite likely that 
\textit{Planck}
will turn out to be the end of the line in the generation 
of high-precision satellites.
\textit{Planck} will be able to distinguish between 
different types of inflationary models, and it is the most promising
guide to understand the early universe.
Among its many goals, it aims at unveiling the
microphysics of inflation, understanding what are the high-energy theories
beyond those imprinted in the CMBR fluctuations, 
and learning more about dark energy.
In particular, and focusing on inflation theories, 
it might measure non-zero non-gaussianity, 
which is a key observational feature since it
works as a discriminator of microphysics. 

The main object 
of study in the next four chapters will be the bispectrum, which is the 
lowest-order
non-gaussian statistics component, and therefore potentially easier to observe (than higher-point statistics).

\section{Summary of the outline of the thesis}
In this thesis we explore the inflationary signatures of single-field models, 
and study how the universe would have reheated after a Dirac--Born--Infeld
inflation phase.
One of the key questions we aim to address is: 
how can we efficiently use \textit{Planck}'s data to solve for
 the degeneracy in the inflationary models?
 By this we mean that several, distinct sets of parameters
 in a given theory can combine in such a way so as to produce the same predictions
 for observables. 
 What we would like to know is 
 how to build an efficient dictionary between 
 observations and parameters of a theory.
 
 We start by reviewing non-gaussinities in chapter \ref{chapter:two}, 
 explaining how they encode blueprints of the inflationary models.
 We first focus on $P(X,\phi)$
 theories and compute the amplitude of their bispectrum
 to next-order in slow-roll.
 We later generalise our results to the Horndeski class
 of models, which includes all theories that do not have propagating ghosts.

 For some time now we suspect that 
the theoretical uncertainty in estimating the bispectrum 
 is larger than the experimental errors \cite{Chen:2006nt}.
If we expect to constrain the parameters of the theory using \textit{Planck}'s data,
one necessarily needs to push the theory-error down, otherwise the 
advantages of having high-precision measurements are not met. This can be done 
in much the same way as electroweak precision tests some decades ago,
when physicists realised that they had
to control the theoretical inaccuracy 
of their calculations 
to be able to keep pace 
with the precision of their experiments (see, for
example, the review by Altarelli \etal \, \cite{Altarelli:1997et}). 
In chapter \ref{chapter:three} we give a more accurate estimate of the bispectrum, 
by reducing the maximum 
theory-error by several tens of percent (around 95\% in most cases).
 
 In chapter \ref{chapter:four} we present
 a careful analysis of the scale and shape-dependences 
 of the bispectrum and 
 we elaborate on its discriminatory power 
 to tell inflationary models apart.
 We focus on bispectrum shapes, rather than the amplitude of the 
 bispectrum, to make a most efficient use of 
 empirical evidence of inflationary theories.
In chapter \ref{chapter:two} we will see that 
the bispectrum depends only on a handful of parameters, 
which requires finding an efficient algorithm
to solve for this degeneracy.
We argue that using templates (as commonly done in CMBR data analysis)
might not be the most promising strategy to understand the 
early universe microphysics.

In chapter \ref{chapter:five} we go beyond the slow-roll approximation, and 
ask the following question: what if inflation was not close 
to de Sitter? In that case, one could wonder whether the inflationary
signatures in the CMBR would be different form those obtained and 
discussed in chapters \ref{chapter:two}, \ref{chapter:three} and \ref{chapter:four}. 
We obtain estimates for the bispectrum 
under the assumption that the spectrum of perturbations is approximately
flat, consistent with observational data. 
We comment on the implications for the scale-dependence of 
the bispectrum in these theories, and argue this behaviour can be explored
using complementary observational tests.

Chapter \ref{chapter:six} focuses on the study of preheating after a period of 
Dirac--Born--Infeld inflation. We study how matter particles 
could have been created and efficiently repopulated the universe.
We comment on the implications of the reheating temperature
in such theories.

Finally, in chapter \ref{chapter:seven} 
we review the main findings of this thesis
and point out to possible extensions to future work.
The appendices collect important material for the 
developments of the chapters in the main body of this thesis.

\clearpage

\newpage
\thispagestyle{empty}
{\color{white} kuwhf}
\newpage
\clearpage

\begin{savequote}[30pc]

We choose to go to the moon in this decade and do the other things, not because they are easy, but because they are hard, because that goal will serve to organise and measure the best of our energies and skills, because that challenge is one that we are willing to accept, one we are unwilling to postpone [...]

\qauthor{John F. Kennedy}
\end{savequote}

\chapter{Non-Gaussianity in Inflationland}
\label{chapter:two}

\hrule
\vspace*{1cm}

For a long time theoretical cosmology's only exercise 
was to develop theoretical models which would 
one day be put to scrutiny
against observations as working models of the early universe.
We are living in an exciting era of precision cosmology where, 
perhaps for the first time, we can gather enough data with
the appropriate resolution to test those models. 
This data might not be able to rule out some inflation models completely, 
as some degree of fine-tuning in model building can accommodate even unfavourable data.
Nevertheless, it will certainly put some models under moderately high pressure.
We will ultimately want to know what was the physics behind inflation, 
if inflation did indeed happen. Our best hope to learn about this mechanism
is through non-gaussianities.

\para{Outline} We start by reviewing in \S \ref{sec:evolution_NGs} 
the work that has been done on non-gaussianities,
regarded as the most 
promising discriminator of microphysics.
This chapter therefore contains important review material,
but also original work.
	In~\S\ref{sec:singlefieldinflation} we discuss the background model
	and specify our version of the slow-roll approximation.
	Results involving the lowest powers in slow-roll parameters will 
	be called ``leading-order,'' whereas those involving one 
	extra power in slow-roll will be referred to as 
	``next-order,'' and so forth.
	Our initial discussion will be focused on $P(X,\phi)$ models.
	In~\S\ref{subsec:second} we recapitulate the computation of the
 two-point function of perturbations (including its explicit scale-dependence).
 We discuss the derivation of the third-order action
 for the primordial perturbation in~\S\ref{sec:third},
	and based on recent developments, we extend our calculation to 
	\textit{all} single-field models which \textit{do not} contain ghost-like instabilities.
	We give in \S\ref{subsec:inin} a brief description of
	 the ``in--in'' (or Schwinger--Keldysh) formulation for
	expectation values, necessary to compute correlation functions.

This chapter includes material based on work done in collaboration with 
Clare Burrage and David Seery, published in Ref. \cite{Burrage:2011hd}.

	\section{Understanding non-gaussianities}
	\label{sec:evolution_NGs}

As we discussed in the introduction to this thesis, 
our picture of the early universe has become much more 
precise with access to data with unparalleled sensitivity.
This has been made possible with the advent of high-precision 
satellites, like \textit{WMAP} and more recently \textit{Planck}. 
If the temperature in the microwave sky was \textit{perfectly} 
homogeneous, there would be little to know about the 
CMB, except that its spectrum was that of blackbody radiation.
Fortunately this is not the case. 
Below is the microwave sky as seen by the \textit{WMAP} satellite.

\begin{figure}[htpb]
\begin{center}
\includegraphics[width=13cm]{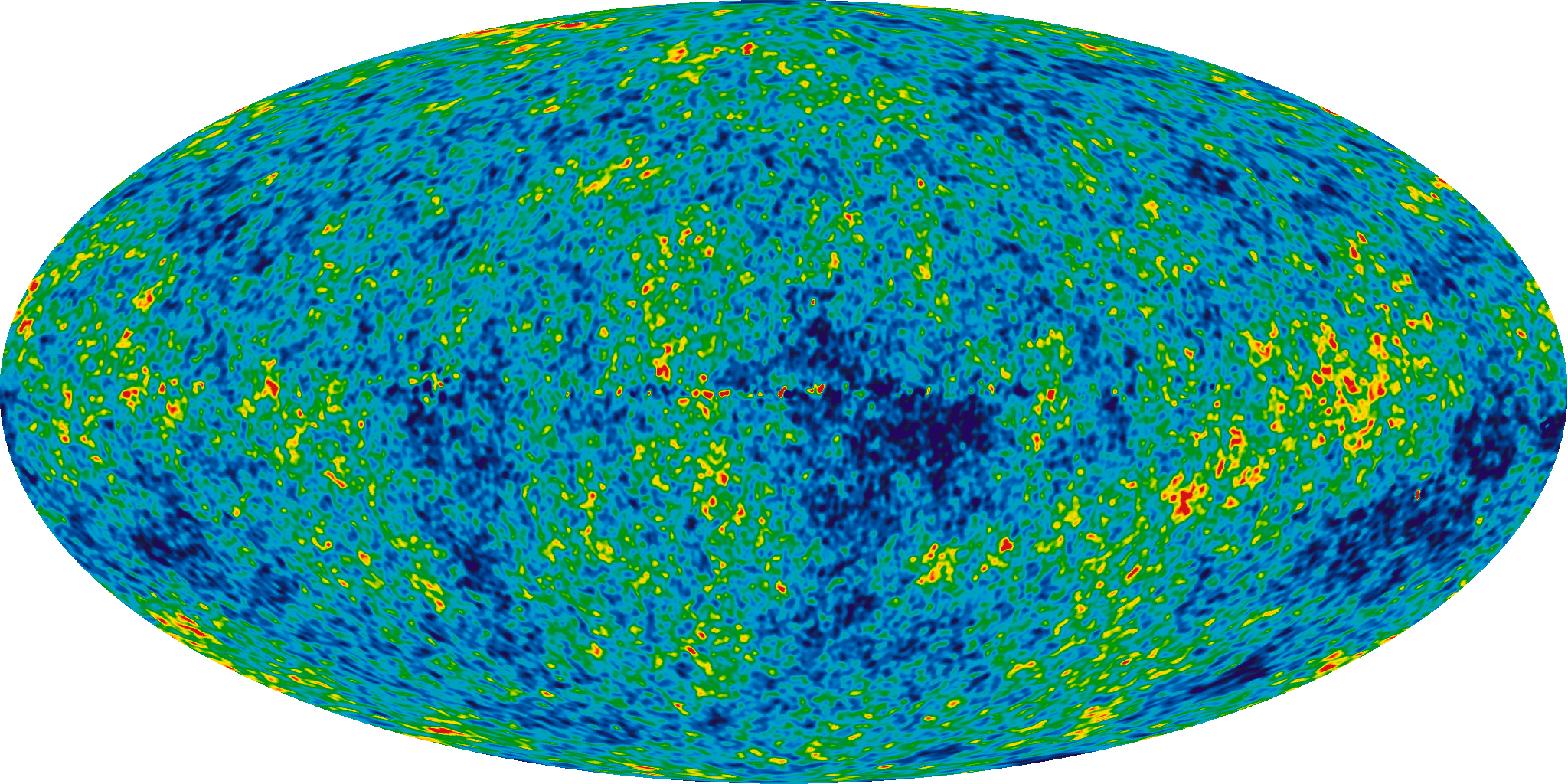}
\label{fig:wmap-chapter2}
\caption{Detailed map of the microwave sky as seen by \textit{WMAP}. 
This is a snapshot of the universe when it was only 380,000 years old.}
\end{center}
\end{figure}

The colours in this picture not 
only say that the microwave sky
does not have the same temperature everywhere.
They mainly tell us that there is a wealth of information
encoded in the CMBR which is sensitive to the microphysics
during inflation.
It is in the way that the temperature in one point 
correlates with that in other points that 
we can learn about the inflationary mechanism.
But how?

The CMBR is usually thought to have its origins in a 
primordial era \cite{Komatsu:2009kd} governed by 
quantum fluctuations. For a review on the numerous 
mechanisms for generation of quantum fluctuations during inflation
see, for example, the book by Lyth \& Liddle
		\cite{Lyth:2009zz}.
During inflation, cosmological scales were pushed outside the 
observable horizon. At the same time, quantum fluctuations 
were allowed to grow to later become classical far outside the horizon.
	In its simplest implementation, inflation forecasts
	an approximately scale-invariant, Gaussian distribution of
	perturbations \cite{Bardeen:1983qw, Wands:2000dp}.
	This means that fluctuations are only slightly stronger 
	on larger scales (since the spectrum is red-tilted),
	and they are completely characterised by their two-point correlations
	(the power spectrum of perturbations). 
	Although minimalist, these predictions are in good agreement with
	present-day observations, including \textit{WMAP} 7-year data
	\cite{Jarosik:2010iu, Larson:2010gs, Komatsu:2010fb}.
	
	Given the recent improvement and sophistication of CMB experiments, 
	we might now be able to detect
	non-zero three- and higher
	$n$-point correlations \cite{Verde:1999ij,Komatsu:2001rj}.
	These are globally known as \textit{non-gaussianities}
	and they can be thought of as including important information
	about interactions within the inflaton sector, and others, during inflation.
	They are an intricate product of whatever inflationary process
	occurred in the early universe. 
	Understanding and identifying the non-gaussian signatures 
	in each inflationary model is comparable to producing an
	inflationary fingerprint,
	which can be later identified in the microwave sky.
	On the other hand, other constraints will soon emerge 
 from the non-gaussian statistics of collapsed
	objects, using LSS data \cite{Bernardeau:2001qr}.
	We therefore expect non-gaussianity constraints to arrive from 
	different sets of data.
	
	In principle, valuable information is encoded in each
	$n$-point function. In practice, extracting information from
	the four-point function is 
	computationally challenging, as described in Refs.
	\cite{Smidt:2010ra,Fergusson:2010gn};
	it is also unclear whether data constraints 
associated with higher $n$-point functions	
	can ever be used efficiently as model discriminators.
	For this reason, attention
	has focused on the three-point correlations, 
	also known as the \textit{bispectrum}. 
	These are most usually cited in terms of a rescaled 
	amplitude, $\fNL$
	\cite{Verde:1999ij,Komatsu:2001rj}, although as we shall see
	in this thesis, there is a lot more information on the bispectrum, 
	which can and should be used to draw constraints.\footnote{The definition
										of $\fNL$ will be presented in
										\S\ref{sec:bispectrum}. 
										For now it suffices to know that
										$\fNL$ is a rescaled magnitude of the 												bispectrum.}
	For a Gaussian field there is no extra information on higher $n$-point correlators, 
	besides the one provided by the power spectrum,
	and $\fNL = 0$. This is the least desirable situation, which is 
	also not excluded from present-day data. 

	Our work assumes there was only one active single scalar field, $\phi$, during inflation
	and that its quantum fluctuations, $\delta \phi$, became imprinted in the CMB. 
	The mapping we want to establish is that between a microphysical Lagrangian
	of perturbations and
	its correlation functions, $\langle \delta \phi^3 \rangle$.
	Early references  are the works by 
	Starobinsky \cite{Starobinsky:1986fxa}, 
	Sasaki \& Stewart \cite{Sasaki:1995aw}
	and
	Lyth \& Rodr\'{i}guez
	\cite{Lyth:2005fi}. 
	We now review some of the notable developments
	in this area.
	
	\subsection{Historical developments}
	The first calculations of correlators of 
	quantum fluctuations $\delta \phi$
	were presented shortly after the
	inflationary paradigm was proposed
	\cite{Bardeen:1983qw,Guth:1982ec,Hawking:1982cz,Hawking:1982my,
	Lyth:1984gv,
	Mukhanov:1985rz,Sasaki:1986hm}.
Results were obtained for 
	the two-point functions of single-field 
	inflation theories with canonical kinetic terms.
	
The possibility of producing non-gaussian 
fluctuations during inflation was first investigated 
by Allen, Grinstein \& Wise
in 1987 \cite{Allen:1987vq}. The
first calculation of the three-point correlator
of the temperature anisotropy
dates as early as 1992 by Falk, Rangarajan \& Srednicki
	\cite{Falk:1992sf},
	followed by other partial results 
	\cite{Gangui:1993tt,Wang:1999vf}.
	The complete three-point function was laid down
	in 2002 by Maldacena \cite{Maldacena:2002vr}, in a paper which set 
	up all the subsequent calculations of the bispectrum.
	Three-point correlators originated from multiple fields
	were originally calculated by Seery \& Lidsey
	\cite{Seery:2005gb}. Calculations of four-point correlations
	first
	appeared in Ref. \cite{Seery:2006vu} by Seery, Lidsey \& Sloth,
	and Ref. \cite{Seery:2008ax} by Seery, Sloth \& Vernizzi.
	
	Maldacena's epic calculation 
showed 	
	 that $\fNL$ would be unobservably
	small in single-field models with canonical kinetic terms. 
	More precisely, $\fNL \sim r $, where $r$ is the tensor-to-scalar
	ratio, constrained by observation to
	satisfy $r \lesssim 0.2$
	\cite{Komatsu:2010fb}.
Soon cosmologists turned their attention to more complicated realizations
of single-field theories of inflation.	
Two classes of possibilities emerged:
in effective field theories 
(see, for example, Ref. \cite{Kaloper:2002uj}) Lagrangians naturally contained 
high-order operators of derivatives of the scalar field, suppressed
by some high-energy scale; in other
theories the scalar field $\phi$ was not canonically normalised.

	The first class of these was studied by Creminelli in 
	2003 \cite{Creminelli:2003iq}, who concluded
	that if the dominant kinetic operator
	for the slowly rolling background field was of the form
	$(\partial  \phi)^2$, then
	the three-point correlations of its perturbations
	were effectively indistinguishable
	from Maldacena's simplest model
	\cite{Creminelli:2003iq}. 
	This analysis relied on the slow-roll approximation. 
	The result was that 
	non-gaussianities in these models would disappointingly have 
	negligible amplitude, and could not be used as a 
	diagnostic tool.
	
	\para{Non-canonical models}In theories where other operators than $(\partial  \phi)^2$ 
	are dominant, however, 
	$\fNL$
	can become quite significant, depending on the precise 
	form of the Lagrangian of the background theory.
Examples of these more complicated scenarios are
	``ghost inflation'' \cite{ArkaniHamed:2003uy, ArkaniHamed:2003uz},
	and Dirac--Born--Infeld (DBI) action \cite{Alishahiha:2004eh}.
	In these models 
	the scalar field action governing the dynamics
	of both background and perturbations can be
	written as
	\begin{equation}
		S=\dfrac{1}{2} \int \d^4 x \;
		\sqrt{-g} \ \Big\{R+ 2 P(X,\phi) \Big\} \ \ ,
		\label{eq:startingaction}
	\end{equation}
	where $X \equiv - (\partial \phi)^2$.
	The first term is the Einstein-Hilbert action 
	involving the space-time Ricci scalar $R$,
	and $P(X,\phi)$ is an arbitrary function of $\phi$
	and its first-derivatives through $X$.
	Because of the arbitrariness of $P$, these models
	can accommodate non-canonical kinetic structures.
	
	In ghost-inflation, the kinetic term of $\phi$ 
	has the wrong sign and the action contains 
	several $P(X,\phi)$ contributions among higher-order
	derivatives of $\phi$, as follows
	\[ S=\int \d^4x \sqrt{-g} \bigg\{
	M^4 P_1(X) +M^2 P_2(X) (\Box \phi)^2+
	M^2 P_3(X) \partial^{\mu} \partial^{\nu} \phi \partial_\mu \partial_\nu \phi +\cdots 
	\bigg\} \ \ , \]
	where $P_1$, $P_2$ and $P_3$ are dimensionless 
	functions of $X$ (which is also dimensionless), 
	$M$ is some mass-scale and 
	$\cdots$ represent higher derivative operators acting on $\phi$.
Since this action contains operators with more than one derivative 
acting on the scalar
field, it gives rise to equations of motion which contain at least third-order 
derivatives of $\phi$, signalling a ghost instability.
Arkani-Hamed \etal \, showed that 
the solution of interest takes the form
\[ \phi= c t \ \ , \]
	where $c$ is some dimensionless constant. They also showed 
	that this solution
	protects the theory against radiative corrections, even 
	though the action contains non-renormalisable operators.
	
	In contrast, in DBI inflation it is the presence of a 
	higher-dimensional boost 
	that protects the coefficients of the single derivative 
	operators from large renormalisations.
	The action for the background field in this model takes the form 
\[ 		P(X, \phi)
 		=
 		-\dfrac{1}{f(\phi)}
 		\left\{
 			\sqrt{1-f(\phi) \ X} -1
 		\right\}
 		-
 		V(\phi)\ \ ,
	\]
	where $f(\phi)$ has units of $[\textrm{mass}]^{-4}$.
	The kinetic structure of $\phi$ is embedded in the non-analytic square root, 
	which allows to sum an infinite number of powers of 
	single derivative operators.
	We will review this inflation model in \S \ref{sec:dbi}, 
	and again in \S\ref{sec:dbi-dynamics} from a brane-world
	perspective.
	
Related models based on
	``galileon'' actions have also been obtained
	\cite{Burrage:2010cu,Kobayashi:2010cm,Mizuno:2010ag,Creminelli:2010qf}.
		``One of the reasons why these models raised so much interest
	was because they might leave fingerprints in the CMBR \cite{Vazquez:2008wb,Gao:2010um}. "
	A galileon singlet owes its name 
	to invariance under the transformation
	\begin{equation}
		\phi(x) \rightarrow  \phi(x) + b_{\mu}x^{\mu} + c ,
		\label{eq:galileon-xfm}
	\end{equation}
	for constant $b_\mu$ and $c$,
	under which gradients of $\phi$ are shifted by a constant.
	Eq.~\eqref{eq:galileon-xfm} is a space-time version of
	a galilean transformation,
	first noticed in the Dvali--Gabadadze--Porrati (DGP) model
	\cite{Dvali:2000hr, Deffayet:2001pu}.
	
	Galileons have recently become popular
	since they can be interpreted as
longitudinal graviton modes
	near the decoupling limit
	of massive gravity,
	when $\Mp \rightarrow \infty$,
	while the cutoff of the theory remains fixed	
	 \cite{deRham:2010ik,deRham:2010kj,Hassan:2011hr}.
	 From Eq. \eqref{eq:galileon-xfm} we see that
	the galilean symmetry  
	contains the shift symmetry $\phi \rightarrow \phi + c$, 
	which means that if the background supports a de Sitter 
	solution, then inflation can last for many e-folds, 
	which can be problematic.
	For inflation to last 60 e-folds,  
	the shift symmetry needs to be broken (even if only mildly).
	This is typically achieved by adding a potential.
	However, introducing a potential such as $V \sim m^2 \phi^2$
	will manifestly break the galilean symmetry we started with.
	The absence of this special symmetry implies that the Lagrangian
	of the theory is no longer protected from 
	other galilean violating terms generated via radiative corrections.
	As a result, one looses the motivation of starting with a Lagrangian which
	obeys the galilean symmetry in the first place.
	Nevertheless, it was shown by Burrage \etal \ \cite{Burrage:2010cu} 
	that a non-renormalisation
	theorem still operates, making the dangerous radiative corrections
	small. 
	Therefore, despite these potential problems, 
	galileon inflation sits, alongside DBI inflation,
	as one of the few known examples of a radiatively stable theory of inflation.

	More recently a number of authors have rather
focused on the much milder requirement that 
the Lagrangian preserved unitarity.
The result would be a
sensible quantum field theory
\cite{Deffayet:2010qz, Kobayashi:2010cm, Mizuno:2010ag, Kobayashi:2011nu, Kobayashi:2011pc, Gao:2011qe}, 
where the Lagrangian operators give rise to equations of motion 
 which are at most second-order in derivatives of the field.
In these theories the galilean symmetry is lost
\cite{Deffayet:2009wt} because it can no longer be realised in 
a generic space-time.
These theories go often by the name ``$G$--inflation,'' although as 
we shall see in the end of this chapter, they can also be called 
``Horndeski'' models.

	Models in which the dominant operator 
	of the background kinetic structure is
	different from $(\partial  \phi)^2$ 
	generically fall under the name of 
	\emph{non-canonical} theories.
	Eq.~\eqref{eq:startingaction} was first
	suggested in 1999 by Armend\'{a}riz-Pic\'{o}n, 
	Damour \& Mukhanov
	\cite{ArmendarizPicon:1999rj},
	who named it ``$k$-inflation'' when applied to inflationary cosmologies. 
	The corresponding two-point
	function for scalar perturbations 
	was obtained in the same year by Garriga \& Mukhanov
	\cite{Garriga:1999vw}.
	Three-point correlations induced by~\eqref{eq:startingaction}
	were investigated in 2004 by Gruzinov
	in a decoupling limit where the 
	mixing with gravity fluctuations could be neglected
	\cite{Gruzinov:2004jx};
	gravitational interactions were included in the following
	year by Seery \& Lidsey
	~\cite{Seery:2005wm}.
	The bispectrum of these theories was given in generality by Chen {\etal}
	\cite{Chen:2006nt},
	followed by extensions to multiple fields
	\cite{Langlois:2008mn, Langlois:2008qf, Arroja:2008yy}.
	The four-point function was also computed by a number of authors
	\cite{Huang:2006eha, Arroja:2008ga, Langlois:2008wt,
	Arroja:2009pd, Mizuno:2009mv, Gao:2010xk}.

	\subsection{Non-gaussianities in slow-roll inflation}	
	\label{subsec:estimatesfNL}
	In this chapter we revisit actions of the form~\eqref{eq:startingaction}
	and reconsider the three-point function
	under the slow-roll approximation.
	We do this for two reasons: 
	first, we review our present understanding of
	the bispectrum in $P(X,\phi)$ theories;
	second, we ask whether the estimates described in the literature
	are precise enough to be comparable with the high-sensitivity 
	data soon to be delivered by \textit{Planck}.
	
	The analyses of the $n$-point correlators  
	mentioned above assumed some sort of \textit{slow-roll approximation}
 to control their calculations, usually by restricting
	their results to lowest powers
	of $\varepsilon \equiv - \dot{H}/H^2$
	(or other quantities of similar magnitude, 
	such as $\eta\equiv \dot{\varepsilon}/\varepsilon H$).
	We generically expect this procedure to yield
	estimates accurate to a fractional error of
	order $\varepsilon$,
	which in some models could be as large as $10^{-1}$ to $10^{-2}$.\footnote{In 
													Ref. \cite{Agarwal:2011wm}
													(see also Ref. \cite{Dias:2012nf}) 
													the authors have studied 
													the most likely values for $\varepsilon$ and 
													$\eta$ in the context of brane inflation models.
													There $\varepsilon \lesssim 10^{-13}$. 
													We will assume
													$\varepsilon$ can be as large as $10^{-2}$		
													 for the 
													purposes of illustrating how large the slow-roll
													corrections to the bispectrum can become.
						}
	Whenever $\fNL$ is small, as Maldacena showed in canonical models,
	this is indeed a very good approximation.

	However,
	in models where $\fNL$ is numerically large this
	might not be the case---a fractional error of order $\varepsilon$
	may be
	comparable to
	the
	sensitivity of \emph{Planck}'s data,
	especially if the $\Or(\varepsilon)$ terms enter with
	a relatively large coefficient.
	We will show this is generically what happens in $P(X,\phi)$ theories.
	In the equilateral configuration\footnote{We will be more precise about 
											what we mean by equilateral configuration or 
											template in chapter \ref{chapter:four}.},
	\emph{Planck} will measure $\fNL$
	with an error bar $\Delta \fNL \in [25-30]$;
	future
	CMB experiments such as \emph{CMBPol} or \emph{CoRE} may
	even achieve
	$\Delta \fNL \approx  10$
	\cite{Baumann:2008aq,Core:2011ck}.
	Ideally, we would like the theoretical uncertainty
	in our predictions to fall \textit{below} this threshold.

\para{Next-order corrections}In comparison to the bispectrum of perturbations, 
	corrections to the power spectrum at subleading order in
	$\varepsilon$ are well-understood.
	Stewart \& Lyth
	\cite{Stewart:1993bc} were the first to obtain the propagator for 
	scalar fluctuations up
	to second-order terms in the slow-roll approximation.%
		\footnote{Stewart \& Lyth were obliged to assume that the
		slow-roll parameter $\varepsilon = -\dot{H}/H^2$ was small, 
		thereby truncating the slow-roll approximation accordingly.
		On the other hand, Grivell \& Liddle \cite{Grivell:1996sr} 
		dropped this assumption
		but were unable to obtain analytic solutions.
		Their numerical results confirmed that the Stewart--Lyth
		formulae were valid within a small fractional error.}
 Gong \& Stewart \cite{Gong:2001he,Gong:2002cx} obtained results
 valid to cubic order in $\varepsilon$.
 In what follows we apply the notation 
 introduced by Lidsey \etal \ in Ref. \cite{Lidsey:1995np}:
 results involving the least powers in slow-roll parameters 
 contribute at \textit{leading-order}, followed by \textit{next-order terms} which
 contain contributions with one extra power in slow-roll, 
 and so forth.
 
 Partial results generalising
	Stewart \& Lyth's calculation
	to non-canonical models
were presented by Wei, Cai \& Wang in Ref. \cite{Wei:2004xx}.  	
	Chen
	 {\etal} 	\cite{Chen:2006nt} 
	 computed the power spectrum 
	 to next-order in slow-roll.
	How relevant are the slow-roll corrections to the bispectrum?
	Chen
	 {\etal} 	\cite{Chen:2006nt} calculated the bispectrum to 
	 leading-order in the slow-roll approximation.
	  Next-order contributions to the bispectrum 
	 were identified and presented in terms of quadratures,
	 which made it hard to evaluate the corrections
	 to $\fNL$ at next-order in slow-roll. 
	 
	 Nevertheless, 
	the cubic action for perturbations
	derived by Seery \& Lidsey \cite{Seery:2005wm}
	was perturbative in the amplitude of fluctuations
	around the background field,
	but \textit{exact}
	in slow-roll quantities. The coefficients
	of each operator in the action 
	were shown to be slow-roll suppressed,	
	but the derivation did \textit{not} rely on the slow-roll
	approximation.
It follows that to get an estimate of how much  
next-order terms can change	the amplitude $\fNL$, 
one can evaluate a subset of such corrections
present in the coefficients of the cubic operators.
If these corrections are to be representative of 
	the typical magnitude of next-order terms, 
	then this estimate can be trusted.

What do we find?
	In the case of DBI inflation
	\cite{Silverstein:2003hf,Alishahiha:2004eh},
	the next-order corrections
	entering the coefficients of the cubic operators
	in the action for perturbations,
	generates a fractional correction of
	$101 \varepsilon / 7 \simeq 14 \varepsilon$
	in the equilateral limit. How large can this be?
	For $\varepsilon \simeq 1/20$ (suggested by
	Alishahiha, Silverstein \& Tong \cite{Alishahiha:2004eh})
	this can be of order 70\%.%
		\footnote{Baumann \& McAllister later suggested that
		the Lyth bound \cite{Lyth:1996im} placed a limit on $\varepsilon$
		\cite{Baumann:2006cd}.
		Lidsey \& Huston \cite{Lidsey:2007gq}
		argued that in combination with the
		large value of $\varepsilon$ implied by
		Alishahiha, Silverstein \& Tong \cite{Alishahiha:2004eh}
		(see also Ref.~\cite{Lidsey:2006ia})
		this made the `UV' version of the DBI model microscopically inviable.
		The UV model has other difficulties.
		Bean {\etal} \cite{Bean:2007eh}
		noted that backreaction
		could invalidate
		the probe brane approximation,
		spoiling inflation.
		Moreover, we recall from Refs. 
		\cite{Agarwal:2011wm, Dias:2012nf}
		that in brane inflation scenarios, 
		numerical simulations show that values of $\varepsilon$
		are typically much smaller than one, 
		and so assuming $\varepsilon \sim 10^{-2}$
		can be too optimistic.
		For the purposes of making an estimate
		we are ignoring these details---here, we are interested in
		gauging the impact of these corrections in the results 
		of the bispectrum.}
		This implies that next-order corrections to DBI can be 
		as large as $70$\%. 
		If $\fNL$ is observationally large in the equilateral mode, 
		say between 50 and 250 (in agreement with current bounds), 
		then the corrections can shift $\fNL$ as much as 
		$\Delta \fNL \in [40, 200]$---this window 
		is well within reach of \textit{Planck}'s sensitivity.\footnote{In this estimate
												we are discarding for simplicity
												the sign of $\fNL$.}

	Are there other compelling reasons to evaluate these corrections?
	So far we described how large an effect next-order corrections 
	can have on the \textit{amplitude} of the bispectrum.	
	The bispectrum, however, is a much richer object.
Are there different shapes arising at next-order?
Can they be realised in an inflation model 
without requiring serious fine-tuning? 
If so, the appearance of new shapes could
provide strong evidence in favour of $P(X,\phi)$ theories
controlling the inflationary dynamics. Can we 
use next-order results to learn more about inflation?

	In this chapter we introduce a precise, accurate calculation
	of the slow-roll corrections.
	Our ultimate goal is to resolve 	
	the large theoretical uncertainties
	which do not meet the high-sensitivity standards
	imposed by \textit{Planck}.
	Although next-order calculations are likely to be sufficient
	for \emph{Planck},
	our findings suggest that
	next-next-order results could be in principle
	required by a
	fourth-generation
	satellite such as \emph{CMBPol} or \emph{CoRE}.
	We do not attempt this here.

	\section{Single-field inflation: an overview}
	\label{sec:singlefieldinflation}

	We consider the theory~\eqref{eq:startingaction}
\[ S=\dfrac{1}{2} \int \d^4 x \;
		\sqrt{-g} \ \Big\{R+ 2 P(X,\phi) \Big\}	\ \ , \]
with
	a homogeneous background solution given by the FLRW metric \eqref{eq:flrw2}.
	The corresponding Friedmann equations are
	\begin{equation}
 		3 H^2= 2X P_{,X}-P \quad
 		\textrm{and} \quad
 		2 \dot{H}+3H^2=-P \ \ ,
		\label{eq:friedmannchapter2}
	\end{equation}
	where $P_{,X} \equiv \partial P/\partial X$.
	The nontrivial kinetic structure of $P$
	causes fluctuations of the scalar field $\phi$ to
	propagate with phase velocity, $c_s$,
	different from that of light:	
	\begin{equation}
		\cs^2 
		=
		\dfrac{P_{,X}}{P_{,X}+2XP_{,XX}}\ \  .
		\label{eq:speedofsound}
	\end{equation}
In what follows we shall refer to the phase speed
of fluctuations simply as ``sound speed,'' although 
these are generically different as explained
by Christopherson \& Malik in Ref. \cite{Christopherson:2008ry}.
	In the special case of a canonically normalized field,
	we have $P=X/2-V(\phi)$ and the fluctuations in $\phi$
	propagate at the speed of light.
	We see that in these circumstances, the function $P$
	is actually the pressure of the fluid associated with 
	the scalar field $\phi$ [cf. Eq. \eqref{eq:pressure-chapter1}].

	\subsection{Fluctuations}
	\label{subsec:fluctuations}
	During inflation the universe quickly becomes smooth and isotropic,
	making $\phi$
	spatially homogeneous to a good approximation. At the same time,
	quantum fluctuations generate small perturbations, $\delta \phi$, whose statistics
	we want to calculate. Since $\phi$ is the only scalar field
	in our model, its fluctuations
	must be communicated to the metric. Therefore, for consistency
	we also need to study the metric fluctuations.

	Our freedom to make coordinate redefinitions allows the
	metric and field fluctuations to be studied in a variety of
	gauges \cite{Kodama:1985bj,Malik:2008im}.
	Whatever choice of gauge (corresponding to a threading and slicing of space-time) 
	we should recover the unperturbed 
	FLRW line element
	in the limit of vanishing perturbations.
	
	There is always enough freedom to write 
	the perturbed metric in terms
	of the Arnowitt--Deser--Misner (ADM) metric
	\cite{Arnowitt:1962hi},
	\begin{equation}
		\d s^2 = -N^2 \d t^2 + h_{ij}\, (\d x^i + N^i \d t)(\d x^j + N^j \d t) ,
		\label{eq:adm}
	\end{equation}
	where $N$ is the lapse, $N^i$ is the shift function,
	and  $h^{ij}$ is the intrinsic metric on spacelike hypersurfaces 
	of constant time $t$.
	The ADM formalism is sometimes called the Hamiltonian formulation 
	of General Relativity, which was nicely reviewed by 
	D'Eath in Ref. \cite{D'Eath:1996at}.
	In the absence of perturbations $N=1$ and $N^i=0$, and
	Eq.~\eqref{eq:adm} reduces to the background FLRW metric \eqref{eq:flrw2}, 
	if the spatial metric is flat, $h^{ij}=a^2 (t) \delta^{ij}$.
	The usefulness of this space-time foliation is that $N$ and $N_i$
	do not support propagating modes,
	which are restricted to $\phi$ and the spatial
	metric $h_{ij}$ only. There are three scalar\footnote{These originate 
														from the field perturbation $\delta \phi $, 
														and the scalar components of $h^{ij}$ 
														which are proportional to $a\delta^{ij}$ 
														and $\partial^i \partial^jb$ (any 
														metric can be decomposed into 
														these scalar components, and others).}
	and two tensor modes,
	of which one scalar can be gauged away by choice of spatial
	coordinates and another by choice of time.
	
	At quadratic order
	the tensor modes decouple from the scalar fluctuations.
	This has a simple consequence: 
	when computing the quadratic action for 
	the tensor fluctuations, the scalar field contributions $\delta \phi$
	will not contribute to the answer.
	At tree-level the
	tensor fluctuations only start contributing to the 
	four-point function
	\cite{Arroja:2008ga,Seery:2008ax}.
	The contributions of the tensor modes to the bispectrum
	arise only at loop level, generated
		by insertion of vertices with extra factors of
		the fluctuations. Such diagrammatic expansion is in 
		powers of $H^2/M_P^2 \approx  10^{-10}$, which is
		negligible compared to the contributions from tree-level diagrams.
		In what follows we shall work at tree-level,\footnote{
		Weinberg \cite{Weinberg:2008mc,Weinberg:2005vy,Weinberg:2006ac}
		and van der Meulen \& Smit
		\cite{vanderMeulen:2007ah} have investigated when
		loop diagrams are subdominant compared to tree-level diagrams.}
		and we therefore discard in what follows the tensor modes.
		We will compute in \S\ref{sec:tensor} the spectrum of tensor perturbations.

Regardless of whether or not we take the tensor fluctuations into account, 
we use constraint equations to find the lapse and the shift
as functions of the propagating modes.
This is because $N$ and $N^i$ are, in fact, Lagrange multipliers
in the action. While $N$ is associated with the freedom of choosing 
time reparametrizations, $N^i$ is associated with
spatial coordinates reparametrizations.
Alternatively, we could adopt the path integral formulation for 
		inflationary perturbations proposed in 
		\cite{Prokopec:2010be}, which does not solve for these constraint equations;
		rather the constraints are imposed by introducing 
		auxiliary, non-dynamical fields.
In this thesis, however, we shall adopt the traditional route
and solve for the constraint equations.

The choice of gauge depends essentially on the model
we want to study. If there are multiple fields in the action
then it is common to adopt the \textit{uniform-curvature gauge}, 
whereas in single-field models it is traditional to 
work in the \textit{comoving gauge}.
 In the latter the three-dimensional surfaces 
 are chosen so that $\phi$ is perfectly homogeneous and
 there exists only one
	 propagating scalar mode---this is 
	 absorbed by $\zeta$ in the spatially flat metric $h_{ij}$, via
	\begin{equation}
		h_{ij} = a^2(t) \, \e{2\zeta} \, \delta_{ij} \ \ .
		\label{eq:comoving-curvature-perturbation}
	\end{equation}
	From this equation we read that $\zeta$ is the 
	perturbation of the locally defined scale factor, 
	which encodes the expansion history 
	of the universe  \cite{Bardeen:1983qw, Mukhanov:1990me}. 
	Using the definition of the number of e-folds in 
	Eq. \eqref{eq:efolds}, it follows that 
\begin{equation}
\zeta = \delta N\ \ .
\end{equation}	
This equation is the essence of the ``$\delta N$''
or ``gradient expansion'' formalism, which is 
a popular 
approach for computing the correlation functions of $\zeta$, 
and therefore non-gaussianity---see, 
for example, Refs. \cite{Lyth:2004gb, Lyth:2005fi}.

\para{Notation}At this point we comment on the notation for the 
comoving curvature perturbation used in this thesis,
which agrees with the more recent literature,
		including Refs.~\cite{Salopek:1990jq,Maldacena:2002vr,Chen:2006nt,
		Cheung:2007st,
		Creminelli:2010qf,Baumann:2011su}.
		Historically, 
		the perturbation in the 
	 comoving gauge
was initially denoted by 		
		$\mathcal{R}$---see, for example, the early papers in Refs. 
		\cite{Liddle:1993fq,Lyth:1998xn,Wands:2000dp}.
		The letter $\zeta$ had often been reserved to
		denote the perturbation in the uniform density gauge.
		In either case, in single-field inflation 
		for adiabatic, super-horizon perturbations,
	the comoving and uniform density slicings coincide---up 
	to a(n) (irrelevant) sign convention---as showed by Wands \etal\ \cite{Wands:2000dp}.
	It then follows that for 
	super-horizon evolution (or rather, lack of), we can write
	\[ \zeta = \mathcal{R} \ \ ,\] 	
	and the letters can be interchanged without harm. 
		Throughout this thesis we employ the notation that $\zeta$ is 
		the comoving curvature perturbation.

	\para{Slow-variation parameters}Our first goal is to 
	deduce the action for the small, inhomogeneous perturbations, $\zeta$. 
	It is in this sense that	
	we work perturbatively in $\zeta$.
	This is \textit{independent} of assuming a slow-roll approximation.
	
	As we have discussed in \S \ref{sec:inflationland}, one 
	typically assumes that inflation is described by a de Sitter expansion, when 
	the Hubble parameter, $H$, is constant. 
	To quantitatively describe how far away 
	inflation can occur from a purely de Sitter epoch, 
	one typically introduces slow-roll parameters, 
	as
	defined in Eqs. \eqref{eq:defepsilon} and \eqref{eq:defeta}:
	\begin{equation}
		\varepsilon \equiv - \frac{\d \ln H}{\d N}
		= - \frac{\dot{H}}{H^2} 
		\quad \text{and} \quad 
		\eta \equiv \frac{\d \ln \varepsilon}{\d N}
		= \frac{\dot{\varepsilon}}{H\varepsilon} 
		\quad\ \ .
		\label{eq:slowvariation}
	\end{equation}
	To parametrize the time evolution of the  
	sound speed for perturbations we further introduce
	\begin{equation}
		s \equiv \frac{\d \ln \cs}{\d N}
		= \frac{\csdot}{H \cs} \ \ .
\end{equation}	
	For inflation to occur we expect background quantities to be slowly 
	varying, and to satisfy
\[\varepsilon, |\eta|, |s| \ll 1 \ \ . \]
This is the \textit{slow-roll approximation}.

	\para{Action and constraints}Writing the action \eqref{eq:startingaction}
	in terms of the ADM variables in Eq. \eqref{eq:adm}, 
	we obtain 
	\begin{equation}
		S = \dfrac{1}{2} \int \d^4 x \; \sqrt{h} \,
			N \Big\{
				{R}^{(3)} +2P (X, \phi) 
			\Big\}
			+
			\dfrac{1}{2} \int \d^4 x \; \sqrt{h} \,
			N^{-1} \Big\{
				E_{ij} E^{ij} - E^2
			\Big\}\ \  ,
		\label{eq:admaction}
	\end{equation}
	where $E_{ij}$ satisfies
	\begin{equation}
		E_{ij} = \frac{1}{2} \Big( \dot{h}_{ij} - N_{(i \mid j)} \Big) \ .
	\end{equation}
	Here $\mid$ denotes a covariant derivative with respect to $h_{ij}$,
	and symmetrized indices are enclosed in brackets $(\cdots)$.
	The extrinsic curvature of spatial slices, $K_{ij}$,
	is related to $E_{ij}$ via the shift function, $K_{ij} = N^{-1} E_{ij}$.

The (non-dynamical) constraint equations are 
obtained by varying the action with respect to 	
	 $N$ and $N_i$ \cite{Seery:2005wm}. We find
	\begin{subequations}
	\begin{equation}
		{R}^{(3)} + 2P - 4P_{,X}
			\left( X+h^{ij} \partial_i \phi \partial_j \phi	\right)
		- \dfrac{1}{N^2} \left(E_{ij}E^{ij}-E^2 \right)
		=
		0\ \ ,
		\label{eq:shift-constraint}
	\end{equation}
	and
	\begin{equation}
		\nabla^{j} \bigg[
			\dfrac{1}{N}\left( E_{ij}-Eh_{ij} \right)
		\bigg]
		=
		\dfrac{2 P_{,X}}{N} \bigg(
			\dot{\phi}\partial_i \phi
			- N^{j}\partial_i \phi \partial_j \phi
		\bigg)\ \  .
		\label{eq:lapse-constraint}
	\end{equation}
	\end{subequations}
	Eqs.~\eqref{eq:shift-constraint}--\eqref{eq:lapse-constraint}
	are to be solved order-by-order. To do so, we write
\[ N = 1 + \alpha\ \ , \]	
	where $\alpha$ is some expandable function 
    in powers of the perturbation $\zeta$.
	Likewise,
	the shift vector can be decomposed into its
	irrotational and divergent-free parts,
\[ N_i = \partial_i \theta + \beta_i\ \ , \]	
	where $\beta_i$ satisfies 
	$\partial_i \beta_i = 0$, by assumption.
	We also expand $\theta$ and $\beta_i$  perturbatively
	in powers of $\zeta$, writing the terms of $n^{\mathrm{th}}$ order
	as $\alpha_n$, $\beta_{ni}$ and $\theta_n$.
	As discussed in Refs. \cite{Maldacena:2002vr,Chen:2006nt}, 
	it turns out we only need to solve the
	 constraints to first-order to study the 
	 three-point correlations.\footnote{To be more precise, in general to obtain the 
											$m$-point correlator, we only need to solve the 
											Euler-Lagrange equations to $(m-2)^{\textrm{th}}$-order.					
													}
	To first-order, we find
	\begin{equation}
		\alpha_1 = \dfrac{\dot{\zeta}}{H}\ ,
		\quad
		\beta_{1i} = 0 \ ,
		\quad
		\text{and}
		\quad
		\theta_{1} = - \dfrac{\zeta}{H} + 
			\dfrac{a^2 \SigmaP}{H^2} \partial^{-2} \dot{\zeta} \ \ ,
		\label{eq:order1constraints}
	\end{equation}
	where we have introduced quantities measuring derivatives of $P$
	\cite{Seery:2005wm}:
	\begin{subequations}
	\begin{align}
		\SigmaP
			& \equiv
			X P_{,X} + 2 X^2 P_{,XX}
			=
			\dfrac{\varepsilon H^2}{\cs^2}  \ \ \textrm{and}
			\label{eq:sigmadef}
		\\
		\lambdaP
			& \equiv
			X^2P_{,XX}+\dfrac{2}{3}X^3P_{,XXX} \ \ .
			\label{eq:lambdadef}
	\end{align}
	\end{subequations}
	Some of our formulae will be expressed in terms of these variables.

	\subsection{Two-point correlations}
	\label{subsec:second}
	
	To compute the two-point statistics of the 
	comoving curvature perturbation
	we first need to obtain the 	
	action~\eqref{eq:admaction} to second-order in $\zeta$. This was
	first done for $P(X,\phi)$ theories by Garriga \& Mukhanov
	\cite{Garriga:1999vw}, who have found
	\begin{equation}
		S^{(2)}
		=
		\int \d^3x \, \d\tau \;
		a^2 \zterm
		\big\{
			(\zeta')^2 - \cs^2 (\partial \zeta)^2
		\big\} \ \ ,
		\label{eq:eom2conformaltime}
	\end{equation}
	where $\zterm \equiv \varepsilon / \cs^2$.
	The form of this action is actually quite general, 
	and we will see it applies to all single-field models
	of interest. 
	To make the comparison with extensions of this chapter easier, 
	we will leave $z$ explicit in our formulae. 
	Accordingly, we define the 
	corresponding slow-variation
	parameter,
	\begin{equation}
		v \equiv \frac{\d \ln \zterm}{\d N} = \frac{\dot{\zterm}}{H \zterm}
		= \eta - 2 s .
	\label{eq:defv}
	\end{equation}
	The last equality applies to $P(X,\phi)$ models only.
	For more complicated theories, such as galileon inflation, 
	this shall not be the case and $v$ can have a rather 
	complicated expression.
	The only requirement we impose on $z$ is that is must be positive
	so that the fluctuations are not ghost-like, and its first-derivative
	must be well-defined.

	To simplify some intermediate expressions,
	it will be necessary to have an expression for the
	variation $\delta S^{(2)} / \delta \zeta$:
\[ \delta S^{(2)}=\int \d^4 x \, 
\dfrac{\partial \mathcal{L}^{(2)}}{\partial \zeta} \partial \zeta\ \ . \]	
 We find
	\begin{equation}
		\dfrac{\partial S^{(2)}}{\partial \zeta}=
		-2aH \partial^2 \chi -2a \partial^2 \dot{\chi}
		+2\varepsilon a \partial^2 \zeta\ \ ,
		\label{eq:eom2zeta}
	\end{equation} 
	where $\chi$ satisfies
	\begin{equation}
		\partial^2 \chi
		=
		\frac{\varepsilon a^2}{\cs^2} \dot{\zeta} \ \ .
		\label{eq:chieq}
	\end{equation}
	The equation of motion for $\zeta$ 
	follows by setting $\delta S_2 / \delta \zeta = 0$
	in~\eqref{eq:eom2zeta}---we say $\zeta$ is \textit{on-shell}.
	
	\para{Slow-variation approximation}Although Eqs.~\eqref{eq:eom2conformaltime}
	and \eqref{eq:eom2zeta}--\eqref{eq:chieq}
	are exact at linear order in $\zeta$,
	it is not known how to solve the equation of motion~\eqref{eq:eom2zeta}
	for arbitrary backgrounds in conformal time variables, 
	when $\{ \varepsilon , \eta, s \}$ might be non-perturbative.
	We will show how this can be done using 
	different time coordinates in \S\ref{sec:dynamics}.
	
	As an alternative approach, Lidsey {\etal} \cite{Lidsey:1995np}
	noted that the time-derivative of each slow-variation parameter
	is proportional to a sum of products of slow-variation parameters,
	overall contributing at next-order in slow-roll.
Therefore, assuming 
	\begin{equation}
		0 < \varepsilon \ll 1,
		\quad
		|\eta| \ll 1,
		\quad
		|s| \ll 1
		\quad
		\text{and}
		\quad
		|v| \ll 1 ,
		\label{eq:slow-variation-approximation}
	\end{equation}
	and working to first-order in these quantities, we may formally treat them as
	constants. The calculation will then be 
	organised in increasing powers of these quantities. 
	Corrections contributing at or higher-order 
	than next-to-next order 
	will not be kept, since we expect them to be strongly subdominant.
	Indeed, in general we expect next-to-next order terms to be suppressed
	compared to next-order corrections by the same amount that next-order terms
	are corrections to leading-order results.
	
	The expansion of a general background quantity 
	under the slow-variation approximation above 
	works as follows.	
	Take the Hubble parameter, $H(t)$, to be 
	a representative background quantity.
	Expanding this variable up to next-order in slow-roll
	around a reference time, $t_\star$, yields
	\begin{equation}
		H(t) \simeq H(t_\star) \left\{ 1 + \varepsilon_{\star}
			\Delta N_\star(t) + \cdots \right\} ,
		\label{eq:time-dependence}
	\end{equation}
	where $\Delta N_\star(t) = N(t) - N(t_\star)$ denotes 
	the number of elapsed e-folds since the reference time.
	The reference time is just a pivot scale and we anticipate 
	that physical quantities (that is, \textit{observables}) cannot depend
	on the arbitrary reference scale $t_\star$, or
	equivalently $\Delta N_\star$.
	This is precisely equivalent to what happens in renormalisation techniques in 
	quantum field theory.
	We expect 
		Eq.~\eqref{eq:time-dependence}
	to yield a good approximation to the full time evolution whenever
	$|\varepsilon_\star \Delta N_\star(t) | \ll 1$
	\cite{Gong:2001he,Gong:2002cx,Choe:2004zg}.
	We see that this approximation fails	
	when $\Delta N_\star(t) \sim 1/\varepsilon_\star$; also, if some of the slow-variation 
	parameters become (even) temporarily large around the time of
	horizon crossing, as in ``feature models''
		\cite{Chen:2006xjb,Chen:2008wn,Chen:2010bka,Leblond:2010yq,Adshead:2011bw}, 
		then this approximation breaks down.
	It follows that the approximation \eqref{eq:time-dependence}
	can only be trusted a few e-folds after the reference scale has 
	exited the horizon.

	How can we then study the super-horizon evolution 
	of the perturbations if the approximation \eqref{eq:time-dependence}
does not seem to apply in this asymptotic limit?	
The super-horizon limit corresponds to 
	\textit{many} e-folds after horizon crossing, when typically
	one needs to apply an
	improved formulation of perturbation theory
	obtained by resumming powers of $\Delta N$
	\cite{Seery:2007wf,Burgess:2009bs,Seery:2010kh}
	(various such formalisms are in use
	\cite{Lyth:2005fi,Mulryne:2009kh,Mulryne:2010rp,Peterson:2010mv, Seery:2012vj}).
	This is usually the case in multi-field inflation, when
	ordinary perturbation theory breaks down because 
	of large $\varepsilon$ and $\eta$, and 
	one needs to invoke some sort of renormalisation technique.
In single-field inflation, however, this difficulty 
does not arise---this is because on super-horizon scales, when 
spatial gradients can safely be neglected, 
$\zeta$ is conserved \cite{Rigopoulos:2003ak,Lyth:2004gb,Naruko:2011zk}.
We expect that the same conservation theorem applies to the 	
correlation functions of $\zeta$.\footnote{We will show that the three-point correlator
											is time-independent on super-horizon scales
											for all the single-field models studied in this thesis.}
We will comment on the time-independence of the 
two-point correlator shortly.
		
	\para{Two-point function}The time-ordered two-point function is
	the Feynman propagator,
\[\langle \timeorder \, \zeta(\tau, \vect{x}_1) \zeta(\tau', \vect{x}_2)
	\rangle = G(\tau, \tau'; |\vec{x}_1 - \vec{x}_2|) \ \ , \]	
	which depends on the 3-dimensional invariant
	$|\vec{x}_1 - \vec{x}_2|$.
	Moving to Fourier space variables $G = \int \d^3 q \, (2\pi)^{-3} G_q(\tau, \tau')
	\e{\im \vec{q} \cdot ( \vec{x}_1 - \vec{x}_2 )}$,
	we can write
	\begin{equation}
		\langle
			\timeorder \,
			\zeta(\vec{k}_1, \tau)
			\zeta(\vec{k}_2, \tau')
		\rangle
		=
		(2\pi)^3
		\delta(\vec{k}_1+\vec{k}_2) \,
		G_k(\tau, \tau') \ \ .
	\end{equation}
	The $\delta$-distribution enforces conservation of
	three-momentum, and therefore
	$k = |\vec{k}_1| = |\vec{k}_2|$.
	Breaking up the propagator into 
	elementary wavefunctions of the primordial perturbation
	\begin{equation}
		G_k (\tau, \tau') = \left\{
		\begin{array}{l@{\hspace{5mm}}l}
			\zeta_k(\tau)\, \zeta_k^\ast (\tau')  & \text{if $\tau < \tau'$} \\
			\zeta_k^\ast (\tau) \, \zeta_k(\tau') & \text{if $\tau' < \tau$}
		\end{array}
		\right. .
		\label{eq:propagator}
	\end{equation}
	The elementary wavefunction $\zeta_k$ is a positive frequency solution
	of~\eqref{eq:eom2conformaltime} with 
	the on-shell requirement $\delta S_2 / \delta \zeta = 0$.
	Working to next-order in the 
	slow-variation approximation \eqref{eq:slow-variation-approximation}, 
	we find the time evolution for the perturbation modes 
	\begin{equation}
		\zeta_k(\tau)
		=
		\dfrac{\sqrt{\pi}}{2\sqrt{2}}
		\dfrac{1}{a(\tau)}
		\sqrt{\dfrac{-(1+s) \tau}{\zterm(\tau)}}\
		H_{\frac{3}{2} + \varpi}^{(2)}\left[-k \cs (1+s) \tau \right] \ \ ,
		\label{eq:wavefunction}
	\end{equation}
	with $H_{\nu}^{(2)}$ being the Hankel function of the second kind
	of order $\nu$, and $\varpi \equiv \varepsilon+v/2+3s/2$.
	At sufficiently early times when the modes are well 
	within the horizon, $|k \cs \tau| \gg 1$,
	they cannot feel the curvature of space-time, and 
	Eq. \eqref{eq:wavefunction} describes the time evolution 
	of the perturbations
	in Minkowski space \cite{Bunch:1978yq}.

	\para{Power spectrum}The two-point function at equal times defines
	the power spectrum $P(k,\tau)$ 
	\begin{equation}
		P(k, \tau) = G_k(\tau, \tau) \ \ .\footnote{$P$ here denotes the power spectrum of perturbations
										and should not be confused with the function
										$P(X,\phi)$ used to specify non-canonical theories.}
		\label{eq:power-spectrum-def}
	\end{equation}
	In general, the power spectrum evolves in time, and one needs 
	to specify the time at which it is to be evaluated.
	Using Eq. \eqref{eq:propagator} for $\tau' \rightarrow \tau$,
	working in the super-horizon limit $|k \cs \tau| \rightarrow 0$
	and expanding all the background quantities 
	uniformly around a reference time $\tau_\star$, 
	we can safely neglect the decaying mode and approximate the time evolution
	of the perturbations by the growing mode. We find
	\begin{equation}
		P(k) = \frac{H_\star^2}{4 \zterm_\star \csstar^3}
			\frac{1}{k^3} \left[
				1 + 2 \left\{
					\varpi_\star \bigg(2 - \EulerGamma - \ln \frac{2k}{k_\star}\bigg)
					- \varepsilon_\star - s_\star
				\right\}
			\right]\ \  .
		 \label{eq:powerspectrum}
	\end{equation}
	The Euler--Mascheroni
	constant is
	$\EulerGamma \approx  0.577$.
	We have introduced a quantity
	$k_\star$ satisfying $|k_\star c_{s\star}\tau_\star| = 1$.
	Since to leading-order in slow-roll, 
	$|k_\star c_{s\star} \tau_\star| \simeq
	|k_\star c_{s\star}/ a_\star H_\star|$,
	we describe $\tau_\star$ as the horizon-crossing time associated with
	the wavenumber $k_\star$.\footnote{Chen {\etal} \cite{Chen:2006nt} adopted a definition
										in terms of $|k_\star / a_\star H_\star|$, but the 
										content of their results is identical to ours.}
	The leading-order result in slow-roll is the first coefficient in 
	between the square
	bracket $[ \cdots ]$, and the ``next-order'' correction arises from the
	remaining terms which are one higher-order in the slow-roll expansion.
	Eq. \eqref{eq:powerspectrum} therefore satisfies the organisational 
	scheme of the slow-roll approximation used throughout this thesis.

	The formula for the power spectrum \eqref{eq:powerspectrum}
	allows for a clear statement of the time-independence of the 
	two-point correlator.
	We note that our calculation started with an expansion 
	of \textit{all} the background quantities 
	around some reference time $\tau_\star$, or 
	equivalently, some reference scale $k_\star$.	
	Nevertheless, Eq.~\eqref{eq:powerspectrum}
	does \textit{not} depend on $\Delta N_\star$
	and therefore becomes time-independent
	once the scale $k$ has crossed outside the sound-horizon.
	This is a special property of single-field inflation.
	In classical perturbation theory
	$\zeta$ becomes constant
	and it appears that, in all the examples in the literature, 
	the correlation functions are explicitly time-independent as well.	
	We will return to this issue
	in \S\ref{sec:third-order-action} when we discuss the three-point correlator in detail.

	\para{Scale-dependence}We have observed that the power spectrum 
	\eqref{eq:powerspectrum} is time-independent 
	on super-horizon scales. Moreover it exhibits a 
	weak scale-dependence through the 
	logarithmic term in $P(k)$.
	We conclude that the quantitative predictive power 
	of 
	\eqref{eq:powerspectrum} 
	is limited for scales obeying	
	$|\ln(2k/k_\star)| \lesssim 1$.
	Because $k_\star$ is an arbitrary scale, 
	we always have the freedom to choose $k\sim k_\star$.
	To compute the scaling of the power spectrum, one introduces	
	its ``dimensionless'' version
	$\Ps$ by the rule 
$\Ps = k^3 P(k) / 2\pi^2$.
	We define the spectral index (assuming $k=k_\star$) as
	\begin{equation}
		n_s - 1 = \frac{\d \ln \Ps}{\d \ln k} = -2 \varpi_\star\ \  ,
		\label{eq:rge}
	\end{equation}
	which is valid to lowest-order in slow-roll \cite{Stewart:1993bc}.
	We see that by fixing the scale, one can no longer work with 
	constant background quantities, and 
	their time-dependence needs to be taken into account.
	Alternatively, one could have left the scale-dependence 
	arbitrary, and extracted the coefficients
	multiplying $\ln k$ directly.
	The spectral index is therefore the logarithmic
	scale-dependence 
	of the power spectrum.
	Given the alternative ways of computing the spectral index
	under the slow-roll approximation, 
	it follows that 
	Eq.~\eqref{eq:rge} can be interpreted as
	 a renormalisation group equation 
	 in quantum field theory describing
	the flow of $\Ps$ with $k$, where 
\[\beta_{\Ps} \equiv (n_s - 1) \Ps  \]	
	plays
	the rôle of the $\beta$-function.
	
	The computation of correlation functions of $\zeta$
	is indeed strongly related to renormalisation 
	group techniques popular in quantum field theory.
	The comoving curvature perturbation is related to 
	the field fluctuation via the one-to-one mapping
	\[ \zeta =\dfrac{H}{\dot{\phi}} \delta \phi  \ \ .\]
	It thus follows that
	\begin{equation}
\langle \delta \phi \delta \phi \rangle=
\bigg(\dfrac{H}{\dot{\phi}} \bigg)^{-2}
\ \langle \zeta  \zeta \rangle\ \ ,
\label{eq:correlation-delta-phi}
\end{equation}
	where we have defined 
	$\langle \delta \phi  \delta \phi \rangle= (2\pi)^3 \delta(\vec{k}_1+\vec{k}_2) P_{\delta \phi}(k)$.
	Considering a canonically normalized scalar field and
	using the scalar field equations of motion
	under the slow-roll approximation, we can write
	\[ \bigg( \dfrac{\dot{\phi}}{H}\bigg)^2 
	\simeq 2\varepsilon_\star \bigg\{
	1-\eta_\star \ln (-k_\star \tau)
	\bigg\} \ \ ,\]
	where the second term in brackets captures the next-order corrections.
	Replacing Eq. \eqref{eq:powerspectrum} 
	in Eq. \eqref{eq:correlation-delta-phi}, we find that
	\begin{equation}
\langle \delta \phi \delta \phi \rangle=
\dfrac{H_\star^2}{2k^3 c_{s\star}^2}
\bigg\{
1+2\varpi_\star \bigg[2-\EulerGamma -\ln \bigg( \dfrac{2k}{k_\star}\bigg)\bigg]
-2\varepsilon_\star -2s_\star -\eta_\star \ln (-k_\star \tau)
\bigg\}\ \ .
\end{equation}
Applying the Callan--Symanzik equation
\begin{equation}
\bigg\{
\dfrac{\partial}{\partial \ln k_\star}+\beta_\phi \dfrac{\partial}{\partial \phi}
\bigg\} \ \langle\delta \phi \delta \phi \rangle =0\ \ ,
\end{equation}
with $\beta_\phi \equiv \partial \phi/\partial \ln k_\star$,
we find that 
\begin{equation}
\beta_\phi = -\dfrac{\partial \phi}{\partial N} = \sqrt{2\varepsilon_\star}\ \ ,
\end{equation}
where we have used  $\Delta N = - \ln (-k_\star \tau)$.
We conclude that using renormalisation groups techniques applied to
slow-roll inflation can be used to understand how the 
correlators run with scale.
Furthermore, a generalisation to multi-field inflation can prove useful
in understanding the time and scale-dependences of the correlators.
We do not attempt this here.

	To complete the presentation of the slow-variation catalogue, 
	we define additional slow-variation parameters
	to 
	control the expansion,
	\begin{equation}
		\xi \equiv \frac{\dot{\eta}}{H \eta} \ ,
		\quad
		t \equiv \frac{\dot{s}}{H s}\  ,
		\quad
		\text{and}
		\quad
		w \equiv \frac{\dot{v}}{H v}\ \  .
	\end{equation}
	Table \ref{table:spectral-index} collects the results
	for the spectral index
	valid at leading and next-order 
for different single-field theories.
As argued above, in multi-field models the spectral index will have 
a  super-horizon evolution, and will therefore depend on time---see, for example, Ref. \cite{Dias:2011xy}.

	\begin{table}[h]

	\heavyrulewidth=.08em
	\lightrulewidth=.05em
	\cmidrulewidth=.03em
	\belowrulesep=.65ex
	\belowbottomsep=0pt
	\aboverulesep=.4ex
	\abovetopsep=0pt
	\cmidrulesep=\doublerulesep
	\cmidrulekern=.5em
	\defaultaddspace=.5em
	\renewcommand{\arraystretch}{1.6}

	\begin{center}
		\small
		\begin{tabular}{lll}

			\toprule
		
			model & lowest-order & next-order \\
			\midrule
		
			\rowcolor[gray]{0.9}
				arbitrary &
				$-2\varepsilon_\star - v_\star - 3 s_\star$ &
				$\displaystyle
				-2\varepsilon_\star^2 + \varepsilon_\star \eta_\star
				\bigg(2 - 2 \EulerGamma - 2 \ln \frac{2k}{k_\star}\bigg)
				+ s_\star t_\star
				\bigg( 4 - 3 \EulerGamma - 3 \ln \frac{2k}{k_\star} \bigg)$ \\
			\rowcolor[gray]{0.9}
				& & $\displaystyle
				\mbox{} - 5 \varepsilon_\star s_\star - 3 s_\star^2
				- v_\star( \varepsilon_\star + s_\star)
				+ v_\star w_\star \bigg(
				2 - \EulerGamma - \ln \frac{2k}{k_\star} \bigg)$
				\\[5mm]

				canonical &
				$-2 \varepsilon_\star - \eta_\star$ &
				$\displaystyle
				-2\varepsilon_\star^2 + \varepsilon_\star \eta_\star
				\bigg(1 - 2 \EulerGamma - 2 \ln \frac{2k}{k_\star} \bigg)
				+ \eta_\star \xi_\star
				\bigg( 2 - \EulerGamma - \ln \frac{2k}{k_\star} \bigg)$
				\\[5mm]

			\rowcolor[gray]{0.9}
				$P(X,\phi)$ &
				$-2\varepsilon_\star - \eta_\star - s_\star$ &
				$\displaystyle
				-2\varepsilon_\star^2 + \varepsilon_\star \eta_\star
				\bigg(1 - 2 \EulerGamma - 2 \ln \frac{2k}{k_\star} \bigg)
				- s_\star t_\star
				\bigg( \EulerGamma + \ln \frac{2k}{k_\star} \bigg)$ \\
			\rowcolor[gray]{0.9}
				& & $\displaystyle
				\mbox{} + \eta_\star \xi_\star
				\bigg( 2 - \EulerGamma - \ln \frac{2k}{k_\star} \bigg)
				- s_\star^2 - 3 \varepsilon_\star s_\star - s_\star \eta_\star$ \\

 			\bottomrule
	
		\end{tabular}
	\end{center}
	\caption{$n_s - 1$ at lowest and next-order in the slow-roll approximation.
		The first row applies for arbitrary positive, smooth $\zterm$,
		as explained below Eq.~\eqref{eq:defv}.
		We assume this is the case throughout this thesis.
	\label{table:spectral-index}}
	\end{table}

	\section{Three-point correlations}
	\label{sec:third}
	To compute non-gaussianities, we need the cubic action 
	for $\zeta$. 
	For the action 
	\eqref{eq:startingaction} this calculation was first done
	by Seery \& Lidsey 
	in
	Ref.~\cite{Seery:2005wm}, where a partial result for the three-point
	function was obtained.
	The full three-point function was later obtained by Chen {\etal} 
	\cite{Chen:2006nt}.
	
	\subsection{Third-order action}
	\label{sec:third-order-action}

	We briefly describe the algorithm for computing the cubic 
	action of $\zeta$ here.
	After expanding the action perturbatively in $\zeta$ to 
	third-order, applying numerous integration by parts,
	using Eqs.~\eqref{eq:friedmannchapter2}
	and \eqref{eq:order1constraints},
	we get
	\begin{equation}
		\begin{split}
		S^{(3)} \supseteq 
		& \mbox{}
		\dfrac{1}{2} \int \d^3 x \, \d t \; a^3
		\bigg\{
			- 2 \frac{\varepsilon}{a^2}\,  \zeta (\partial \zeta)^2
			+ 6 \frac{\SigmaP}{H^2}\, \zeta \dot{\zeta}^2
			- 2 \frac{\SigmaP +2\lambdaP}{H^3}\, \dot{\zeta}^3
			- \dfrac{4}{a^4}\, \partial^2 \theta_{1} \partial_j \theta_{1}
				\partial_j \zeta
		\\ &
		\hspace{3.05cm} \mbox{}
			+ \dfrac{1}{a^4}
				\bigg(
					\dfrac{\dot{\zeta}}{H} - 3\zeta
				\bigg)
				\partial^2 \theta_{1}
				\partial^2 \theta_{1}
			- \dfrac{1}{a^4}
				\bigg(
					\dfrac{\dot{\zeta}}{H} - 3\zeta
				\bigg)
				\partial_i \partial_j \theta_{1}
				\partial_i \partial_j \theta_{1}
		\bigg\} \ \ .
		\end{split}
		\label{eq:firstS3}
	\end{equation}
We recall that $\lambda$ and $\Sigma$ were defined 
	in Eqs. \eqref{eq:sigmadef} and \eqref{eq:lambdadef}.	
	In addition we find the 
	boundary contribution to the action:
	\begin{equation}
		S^{(3)} \supseteq  \int_{\partial} \d^3 x \; a^3
		\bigg\{
			-9 H \zeta^3 + \frac{1}{a^2 H} \zeta (\partial \zeta)^2
		\bigg\} \\
		 \ \ .
		\label{eq:boundaryS3}
	\end{equation}

	The action given by the combination of Eqs. 
	\eqref{eq:firstS3} and
	\eqref{eq:boundaryS3} can be further simplified by 
	performing integration by parts. Combining Eqs. 
	\eqref{eq:order1constraints} and \eqref{eq:eom2zeta} 
	with Eq. \eqref{eq:chieq}, one finds%
	\begin{equation}
		\begin{split}
			S^{(3)} \supseteq \mbox{}
			& \frac{1}{2} \int \d^3 x \, \d t \; a^3
			\bigg\{
				\frac{2}{\cs^2 a^2}
					\Big\{
						\varepsilon ( 1 - \cs^2) + \eta \varepsilon
						+ \varepsilon^2 + \varepsilon \eta - 2 \varepsilon s
					\Big\}
					\zeta (\partial \zeta)^2
				\\ & \hspace{3cm} \mbox{}
				+ \frac{1}{\cs^4}
					\Big\{
						6 \varepsilon (\cs^2 - 1) + 2 \varepsilon^2
						- 2 \varepsilon \eta
					\Big\}
					\zeta \dot{\zeta}^2
				\\ & \hspace{3cm} \mbox{}
				+ \frac{1}{H}
					\bigg(
						2 \frac{\varepsilon}{\cs^4} ( 1 - \cs^2 )
						- 4 \frac{\lambdaP}{H^2}
					\bigg)
					\dot{\zeta}^3
				\\ & \hspace{3cm} \mbox{}
				+ \frac{\varepsilon}{2a^4}
					\partial^2 \zeta (\partial \chi)^2
				+ \frac{\varepsilon-4}{a^4}
					\partial^2 \chi
					\partial_j \zeta \partial_j \chi
				+ \frac{2 f}{a^3} \frac{\delta S_2}{\delta \zeta}
			\bigg\}\ \ ,
		\end{split}
		\label{eq:shiftaction}
	\end{equation}
	to which we should add the contributions from the boundary terms
	as follows	
		\begin{equation}
		\begin{split}
			S^{(3)} \supseteq \mbox{}
			& \frac{1}{2} \int_{\partial} \d^3 x \; a^3
			\bigg\{
				- 18 H^3 \zeta^3
				+ \frac{2}{a^2 H} \bigg( 1 - \frac{\varepsilon}{\cs^2} \bigg)
					\zeta (\partial \zeta)^2
				- \frac{1}{2a^4 H^3} \partial^2 \zeta (\partial \zeta)^2
				- \frac{2 \varepsilon}{H \cs^4} \zeta \dot{\zeta}^2
				\\ & \hspace{2.15cm} \mbox{}
				- \frac{1}{a^4 H} \partial^2 \chi
					\partial_j \chi \partial_j \zeta
				- \frac{1}{2a^4 H} \partial^2 \zeta (\partial \chi)^2
				+ \frac{1}{a^4 H^2} \partial^2 \zeta
					\partial_j \chi \partial_j \zeta
				\\ & \hspace{2.15cm} \mbox{}
				+ \frac{1}{2a^4 H^2} \partial^2 \chi (\partial \zeta)^2
			\bigg\}\ \ .
		\end{split}
		\label{eq:shiftactionb}
	\end{equation}	
	In Eq. \eqref{eq:shiftaction} we have defined $f$ to satisfy
	\begin{equation}
		\begin{split}
		f \equiv &
			- \frac{1}{H \cs^2} \zeta \dot{\zeta}^2
			+ \frac{1}{4a^2 H^2} ( \partial \zeta )^2
			- \frac{1}{4 a^2 H^2} \partial_j \zeta \partial_j \chi
			- \frac{1}{4 a^2 H^2} \partial^{-2}
				\big\{
					\partial_i \partial_j ( \partial_i \zeta \partial_j \zeta )
				\big\}
			\\ & \mbox{}
			+ \frac{1}{4a^2 H} \partial^{-2} \partial_j
				\big\{ \partial^2 \zeta \partial_j \chi
					+ \partial^2 \chi \partial_j \zeta
				\big\}\ \ .
		\end{split}
		\label{eq:secondS3}
	\end{equation}
	
On-shell, we observe that the cubic action for perturbations
\eqref{eq:shiftaction}
is given by \emph{five} Lagrangian operators only.
This seems to differ in appearance from the actions obtained in 
Refs.~\cite{Seery:2005wm,Chen:2006nt}. There, 
		a further transformation
		was made to rewrite the terms proportional to 
		the slow-variation parameter $\eta$.
		Using the field equation~\eqref{eq:eom2zeta} and integrating by parts,
		this procedure gives rise to 
		new contributions both to $f$ and to the boundary
		term. Moreover, it also generates
		one extra operator
		$\zeta^2 \dot{\zeta}$, which does \textit{not} appear
		in Eq. \eqref{eq:shiftaction}.
		
		In this thesis we will work with the action 
		 \eqref{eq:shiftaction} for the
		following reasons.
		First, as we shall shortly see, to compute
		the full three-point correlation function, 
		we require the contributions arising from each Lagrangian operator
		separately. Therefore the more operators in the cubic 
		action, the more individual contributions 
		one needs to compute.
		Second, after making the additional transformation,
		the boundary terms will now contribute to the three-point function. 
		To consolidate their contribution one now requires 
		 an appropriate field redefinition, that generates an 
		auxiliary field $\tilde{\zeta}$.
		This redefinition must eventually be reversed to obtain the
		correlation functions of the physical field $\zeta$.
		As we explain below, leaving the action in the form of 
		\eqref{eq:shiftaction} renders such field redefinition unnecessary.

	\para{Boundary terms}The boundary terms defined in a three-dimensional hypersurface 
	in Eqs.~\eqref{eq:boundaryS3} and \eqref{eq:shiftactionb}
	arise in the cubic action from integration by parts with respect to time.
	They were not quoted
	in the original calculation by Maldacena \cite{Maldacena:2002vr},
	neither in subsequent calculations by
	Seery \& Lidsey \cite{Seery:2005wm}, and 
	Chen \etal \ \cite{Chen:2006nt}.
	These calculations explicitly discarded all boundary terms,
	retaining only contributions proportional to $\delta S^{(2)} / \delta \zeta$
	in the action \eqref{eq:shiftaction}.
	The $\delta S^{(2)} / \delta \zeta$ terms were then subtracted by making
	a field redefinition.

	This procedure can be misleading.
	The terms proportional to $\delta S^{(2)} / \delta \zeta$
	give no contribution to any Feynman diagram at any order in
	perturbation theory, because $\delta S^{(2)} / \delta \zeta$ is
	zero by construction when evaluated on a propagator---this is the on-shell evaluation we mentioned before.
	The final answer should therefore be the same
	whether these terms are subtracted or not.
	On the other hand,
	if a field redefinition is performed, 
	then we expect the correlator to be modified accordingly.
	
	Under what conditions is this subtraction procedure correct then?	
	It will yield the correct answer if and only if 
	it reproduces the contribution of the
	 boundary
	component in~\eqref{eq:shiftactionb}.
	This argument was first given in 
	Ref.~\cite{Seery:2005gb},
	and later in more detail in Refs.~\cite{Seery:2006tq,Seery:2010kh},
	but it was
	applied to the third-order action
	for field fluctuations in the spatially flat gauge.
	In this gauge only a few
	integrations by parts are required. The boundary term is
	not complicated and the subtraction procedure works as intended.
	In the present case, however,
	it appears impossible that the subtraction
	procedure could be correct, because the boundary
	action \eqref{eq:shiftactionb} 
	contains operators such as $\zeta^3$ which are not present in
	$f$.
	Indeed, because
	$\zeta$ becomes constant at late times, the $\zeta^3$ term
	seems to diverge
	which should manifest itself as a rapidly evolving
	contribution to the
	three-point function outside the horizon. As we pointed out
	earlier, this is forbidden in single-field inflation.

	This potential problem can be seen most clearly after
	making the redefinition $\zeta \rightarrow \pi - f$
	under which the quadratic action transforms as
	\begin{equation}
		S^{(2)}[\zeta] \rightarrow
		S^{(2)}[\pi]
			- 2 \int_\partial \d^3 x \; a^3 \dfrac{\varepsilon }{\cs^2}
				\dot{\pi} f
			- \int{d^3 x \, \d \tau \; f
				\dfrac{\delta S^{(2)}}{\delta \zeta}}\ \  .
		\label{eq:field-redefinition}
	\end{equation}
	The bulk term proportional to $\delta S^{(2)} / \delta \zeta$ disappears
	by construction.
	After the transformation above, the boundary term becomes
	\begin{equation}
		\begin{split}
		S^{(3)} \supseteq \dfrac{1}{2} \int_{\partial} \d^3 x \; a^3
			\bigg\{ &
				- 18 H \pi^3
				+ \frac{2}{a^2 H} \bigg( 1 - \frac{\varepsilon}{\cs^2} \bigg)
					\pi (\grad \pi)^2
				- \frac{1}{2 a^4 H^3} \partial^2 \pi (\partial \pi)^2
				\\ & \mbox{}
				+ \frac{2 \varepsilon}{H \cs^4} \pi \dot{\pi}^2
				+ \frac{1}{aH} \partial^2 \pi (\partial \chi)^2
				- \frac{1}{a H} \partial_i \partial_j \pi
					\partial_i \chi \partial_j \chi
			\bigg\} \ \ ,
		\end{split}
		\label{eq:boundaryaction}
	\end{equation}
	in which $\chi$ is to be interpreted as a function of $\pi$
	[cf. Eq. (\ref{eq:chieq})].

	The subtraction procedure has produced 
	Eq.~\eqref{eq:boundaryaction} which is not zero.
	This result only leaves the 
	conclusions of
	Refs.~\cite{Maldacena:2002vr,Seery:2005wm,Chen:2006nt} unchanged
	if Eq. \eqref{eq:boundaryaction}
	does \textit{not} contribute to the three-point correlator.
	We will see that this is guaranteed by conservation 
	of $\zeta$ on super-horizon scales.
	Before doing so, we need to briefly recapitulate the 
	 \emph{in-in}
	formalism necessary to compute the correlation functions. 
	We will then return to the issue of
	the contributions to the three-point function from 
	the boundary terms in 
	\S\ref{subsec:boundaryterms}.

	\subsection{Schwinger--Keldysh's \textit{in-in} formalism}
	\label{subsec:inin}

	The correlation functions we want to study are
	equal-time expectation values taken in the state
	corresponding to the vacuum at past infinity.
	They are different from the ``in-out'' calculations
	performed in particle physics which determine the 
	probability of some \textit{in} state (defined at past infinity) 
	becoming an \emph{out} state at future infinity.
	Quantum field theory correlation functions were initially 
studied by Schwinger \cite{Schwinger:1960qe} and 
Keldysh \cite{Keldysh:1964ud} in the 60s, 
and later applied to cosmology by 
Jordan \cite{Jordan:1986ug} and 
Calzetta \& Hu \cite{Calzetta:1986ey}.
But it was only
with Maldacena's epic publication in 2002 \cite{Maldacena:2002vr}
that the applications of the \textit{in-in} formalism
to non-gaussianity were made more clear, 
together with a pair of papers 
by Weinberg \cite{Weinberg:2005vy, Weinberg:2006ac}.
This formalism is the appropriate construction 
to compute expectation values on a time-dependent background, 
and we briefly summarise 
its basic ideas in what follows 
 (there are a number of papers which review the \textit{in-in} formalism---see, for example, Refs. 
\cite{Seery:2006vu, Chen:2010xk, Koyama:2010xj}). 

\vspace*{1cm}

We will be interested in computing three-point correlation functions
at tree-level in the interactions of $\zeta$.

\begin{figwindow}[0,1,\includegraphics[width=6.5cm]
{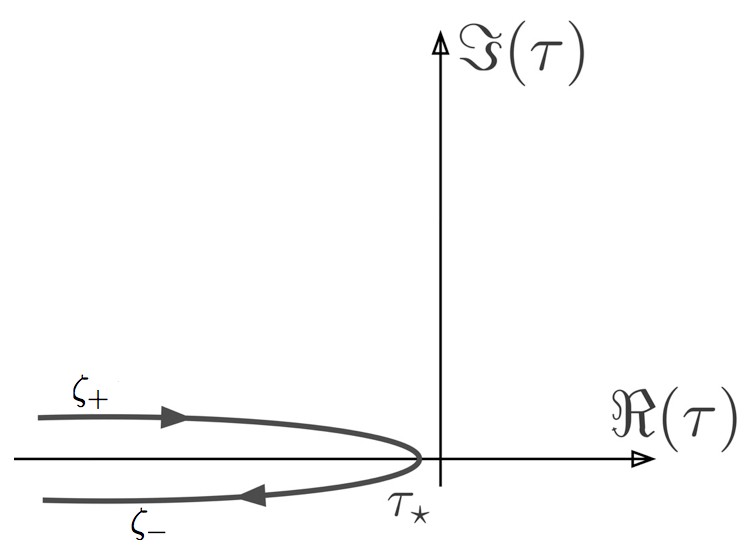}, {Integration contour for the Feynman path integral.}] 
\noindent  	
In terms of the Argand diagram in complex conformal time $\tau$, 
these correlations are obtained by performing a path integral from the true
vacuum of the theory, at $\tau\rightarrow -\infty$, to the time 
of interest when we compute the expectation value, $\tau_\star$.
To this we add the path integral performed backwards, which returns to 
the vacuum at 
$\tau \rightarrow -\infty$. 
\label{fig:argand}
\end{figwindow} 
\vspace*{1.2cm}	
	
\noindent Schematically, this can be translated into the functional integral
\begin{equation}
\langle \zeta (\vec{k}_1) \zeta(\vec{k}_2) \zeta(\vec{k}_3)\rangle
= \int \big[\d \zeta_+ \d \zeta_- \big]\ 
\zeta_+ (\vec{k}_1) \zeta_+(\vec{k}_2) \zeta_+(\vec{k}_3)\ 
e^{\im S[\zeta_+]-\im S[\zeta_-]}\, \delta \big[
\zeta_+ (\tau_\star) - \zeta_- (\tau_\star)
\big]\ \ ,
\label{eq:schwinger}
\end{equation}
where the forward path integral is labelled by the fields $\zeta_+$,
whereas the backwards path integral is labelled by the fields 
$\zeta_-$.
This is \textit{Schwinger's formula}.
In practice, to project onto the true vacuum 
of the interacting theory, one needs to translate the integration contour
to be slightly above and below the negative real axis of conformal time---this is shown in figure 2.2.
This prescription is in many ways similar to the $i\varepsilon$ trick 
recurrent in quantum field theory and ensures the convergence of the 
integral above.
The fields $\zeta_+$ and $\zeta_-$ 
are constrained (by the $\delta$-distribution) 
to agree at any one time \textit{later} than 
that of the observation, so that the path integrals 
are evaluated with an integration contour
necessarily turning 
and crossing the (negative) real $\tau$-axis.
	
	The correlation function can be interpreted in terms of 
	Feynman diagrams: at tree-level and to leading 
	order in the interactions, the three-point correlator is the sum 
	of two Feynman diagrams, depicted in figure 2.3, corresponding
	to the two integrations in Eq. \eqref{eq:schwinger}.
	
	\vspace*{1cm}
	
	\begin{figure}[htpb]
\begin{center}
\includegraphics[width=7cm]{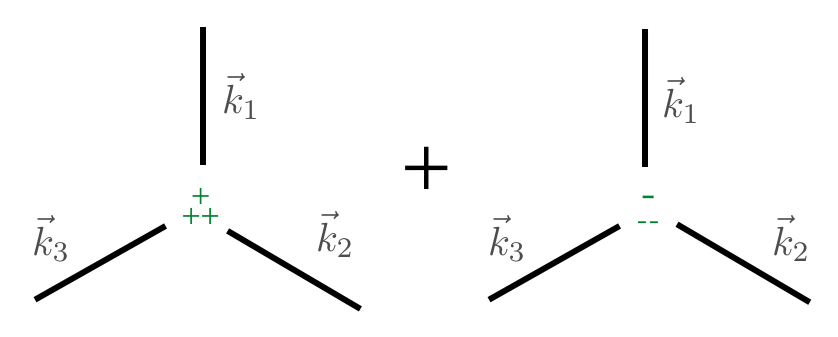}
\label{fig:feynmandiagrams}
\caption{Tree-level Feynman diagrams associated with the 
three-point correlator.}
\end{center}
\end{figure}

\noindent	 Each diagram has three legs with momentum labels $\vec{k}_i$, and at 
	the ``core'' the vertices are evaluated using either the labels 
	$\zeta_+$ or $\zeta_-$.
	The external legs are to be evaluated at a later time, 
	whereas the internal lines are associated to earlier times.
	Each diagram will give a contribution which is precisely the 
	complex conjugate of the other---this means we only need to compute
	\textit{one} of these diagrams for each operator in Eq. \eqref{eq:shiftaction}.
	In total, there will be \emph{five} such calculations to 
	produce the overall bispectrum.
	
	So how are the elementary wavefunctions used to 
	construct the bispectrum?
	To answer this let us first look at an explicit
	calculation of a three-point correlator 
	following Schwinger's formula \eqref{eq:schwinger}.
	Focusing on one Feynman diagram and losing the field label
	to simplify the notation, we write
	\[ \langle \zeta (\vec{k}_1) \zeta(\vec{k}_2) \zeta(\vec{k}_3)\rangle
= \int \big[\d \zeta \big]\  
\zeta (\vec{k}_1) \zeta(\vec{k}_2) \zeta(\vec{k}_3) \ e^{\im S^{(2)}+\im S^{(3)}}\ \ .
\]
Expanding the cubic action of $\zeta$ in the interactions, and 
moving to Fourier space, we get
\begin{equation}
		\begin{aligned}	
		\langle \zeta (\vec{k}_1) \zeta(\vec{k}_2) \zeta(\vec{k}_3)\rangle
= &\int \big[\d \zeta \big]\  
\zeta (\vec{k}_1) \zeta(\vec{k}_2) \zeta(\vec{k}_3) \ e^{\im S^{(2)}}
 \times \\
& \hspace*{1cm} \bigg\{ 
1+\im \int \d^3 x \d \tau  \dfrac{\d^3 q_1\, \d^3 q_2\, \d^3 q_3}{(2\pi)^9} \mathcal{O}\big[\zeta^3 (\vec{q}_i)\big] e^{\im \vec{x} \cdot \sum_i \vec{q}_i} + \cdots
\bigg\}  
\ \ ,
	\end{aligned}
	\end{equation}
where $\cdots$ represent higher-order, slow-roll suppressed contributions, 
and $\mathcal{O}[\zeta^3]$ denotes one of the five cubic 
operators, which contain at least one (time or spatial) differentiated field. 
For the purposes of illustrating how the calculation of the three-point correlator
works though, it suffices to consider $\mathcal{O}[\zeta^3]$
an arbitrary cubic operator in $\zeta$.

The first term in curly brackets offers no contribution to 
the three-point function because it involves an odd number of fields $\zeta$
weighted by a Gaussian measure $e^{\im S^{(2)}}$, which give 
unsuccessful Wick contractions.
Performing the integration in the spatial coordinates $\vec{x}$, 
we are left with
\begin{equation}
		\begin{aligned}	
\langle \zeta (\vec{k}_1) \zeta(\vec{k}_2) \zeta(\vec{k}_3)\rangle
=  \im (2\pi)^3 \int & \d \tau\dfrac{\d^3 q_1\, \d^3 q_2\, \d^3 q_3}{(2\pi)^9}
\delta \bigg(\sum_i \vec{q}_i \bigg) \ \times \\
&\int 
\big[\d \zeta \big]\  
\zeta (\vec{k}_1) \zeta(\vec{k}_2) \zeta(\vec{k}_3) \ e^{\im S^{(2)}}
\mathcal{O}\big[\zeta^3 (\vec{q}_i)\big] + \cdots
\bigg\} 
		\end{aligned}	
\end{equation}
	where the $\delta$-distribution enforces the conservation
	of $3$-momentum.
	
	At this point we invoke a standard, useful result in 
	quantum field theory.
	Let $A$ be a $n\times n$ symmetric, positive-definite matrix.
	Then
	\[ \int \d^n x \ e^{-\frac{1}{2}\, \vec{x}\cdot A \cdot \vec{x}}=
	\dfrac{(2\pi)^{n/2}}{\sqrt{\mathrm{det}\ A}} \ \ . \]
	It follows that for an expectation value, say the two-point correlator
\[ \langle x_i x_j \rangle =A_{ij}^{-1} \ \ , \]	
	where the inverse satisfies $A_{ij}\, A^{jk}=\delta_i^k$.
	Now, for a free field theory
	\[ e^{\im S^{(2)}}= e^{-\frac{1}{2} \int \d z \Phi (z) \, \tilde{Q}\, \Phi(z)   } \ \ , \]
	where $\tilde{Q}$ satisfies $(\Box+m^2)\tilde{Q}=\delta(\vec{z}- \vec{z'} )$, 
	in which $m$ is the mass of the field $\Phi$.
	$\tilde{Q}$ is the inverse of the propagator, 
	and the Green's function of the operator $(\Box+m^2)$.
	We learn that
	\begin{equation}
\int[ \d \zeta_i \d \zeta_j ] \ \zeta_m (\vec{k}_1) \zeta_n (\vec{q}_1)
\ e^{-\frac{1}{2} \int \d z \zeta_i (z) \tilde{Q}_{ij} \zeta_j (z)  }=
\big( \tilde{Q}_{mn}^{-1} \big) (\vec{k}_1, \vec{q}_1)\ \ .
\end{equation}
An integration involving one pair of fields gives rise to a propagator, 
which is built from the elementary wavefunctions as defined in 
Eq. \eqref{eq:propagator}. For each three-point correlator we will require
\emph{three} pairs of \emph{two} wavefunctions, each defined at a different time, 
and the result is 
\begin{equation}
		\begin{aligned}	
\int[ \d \zeta_i \d \zeta_j ] \  \zeta_m (\vec{k}_1) \zeta_n (\vec{q}_1)
\
\zeta_a (\vec{k}_2) \zeta_b (\vec{q}_2)
 &\
\zeta_c (\vec{k}_3) \zeta_d (\vec{q}_3)
\  e^{-\frac{1}{2} \int \d z \zeta_i (z) \tilde{Q}_{ij} \zeta_j (z)  }=\\
&
\big( \tilde{Q}_{mn}^{-1} \big) (\vec{k}_1, \vec{q}_1) \ 
\big( \tilde{Q}_{ab}^{-1} \big) (\vec{k}_2, \vec{q}_2)  \ 
  \big( \tilde{Q}_{cd}^{-1} \big) (\vec{k}_3, \vec{q}_3) 
  \ \ .
  \end{aligned}
\end{equation}
	\noindent Our calculation will be a direct 
	application of this formula.

	\subsection{Boundary terms removed}
	\label{subsec:boundaryterms}

	The boundary terms in Eq. \eqref{eq:boundaryaction}
	were generated from performing integration by parts, 
	using the equations of motion and the field redefinition
	described in Eq. \eqref{eq:field-redefinition}.
	They appear 
	in Eq. \eqref{eq:schwinger}
	as part of the action $S$ with
	support at past infinity and at $\tau = \tau_\ast$.
	The deformed contour of integration
	makes any contribution from past infinity highly suppressed,
	leaving a boundary term evaluated
	precisely at $\tau=\tau_\ast$.
	Because the $\delta$-distribution in Eq. \eqref{eq:schwinger}
	constrains the fields to agree, it follows that 
	any boundary Lagrangian operators 
	not involving \emph{time} derivatives
	produce only a phase
	which cancels
	between the $+$ and $-$ contours.
		This implies that operators involving an admixture 
		of fields and spatial-derivatives of fields 
		give no contribution either.
		We conclude that, in principle, only time-differentiated operators
		can contribute to the 
		three-point function.

	Revisiting the definition of $\chi$ in Eq. \eqref{eq:chieq}, 
	we understand that the entire first line of operators in 	
	\eqref{eq:boundaryaction} does not contribute 
	to the answer we are seeking.
	The argument we have given before that the fields 
	$\zeta_+$ and $\zeta_-$ are constrained to agree at $\tau = \tau_\ast$ 
	(enforced by the $\delta$-distribution), 
	does not apply to their time-derivatives.
		We therefore focus on the last three operators 
	of the boundary action \eqref{eq:boundaryaction}.
	
	Inspection of Eq.~\eqref{eq:boundaryaction} shows that
	the time-derivative operators are of the schematic form
	$\pi \dot{\pi}^2$,
	and therefore lead to a field redefinition of the form
	\[\zeta \rightarrow \pi + \pi \dot{\pi}\ \ .  \]
	This generates operators which have two time-derivatives.
	We now argue that boundary operators with two or more time-derivatives
	are irrelevant on super-horizon scales.
	Using the schematic field redefinition, the three-point correlation
	functions of $\zeta$ and $\pi$ are related by
	$\langle \zeta^3 \rangle
	= \langle \pi^3 \rangle + 3 \langle \pi^2 \rangle \langle \pi \dot{\pi}
	\rangle$
	plus higher-order contributions. However,
	because Eq.~\eqref{eq:powerspectrum}
	implies that the two-point correlator is
	time-independent on super-horizon scales, then
	$\langle \pi \dot{\pi} \rangle \rightarrow 0$
	and therefore we can write $\langle \zeta^3 \rangle = \langle \pi^3 \rangle$,
	up to an irrelevant decaying mode.

	Therefore, on super-horizon scales, the correlation functions of the
	original and redefined fields agree, and after subtraction by a field redefinition,
	the unwanted boundary terms in Eq. \eqref{eq:boundaryaction}
	can be ignored.
	We can also see that the contributions 
	given by time-differentiated operators of order two or higher
	are zero because of conservation of $\zeta$. We conclude that
	time-differentiated operators can be discarded from 
	the three-point functions calculations.

	We observe that the only problematic
	field redefinitions are of the schematic form
	$\zeta \rightarrow \pi + \pi^2$, which arise from boundary
	operators containing a \emph{single} time-derivative.
	Eq.~\eqref{eq:boundaryaction} contains no such operators, 
	and therefore we can discard all the boundary action 
	in what follows.
	We note, however, that this was not guaranteed to be the case, 
	and that a careful analysis of the boundary terms
	must, in general, be performed.
	Shortly before Ref. \cite{Burrage:2011hd} was submitted for publication, 
	a preprint
	by Arroja \& Tanaka appeared \cite{Arroja:2011yj}
	presenting
	arguments regarding the r\^{o}le of boundary terms which are
	equivalent to those of this section.

	\section{Beyond $P(X,\phi)$ and towards Horndeski theories}
	We conclude that the relevant operators in our calculation
	of the three-point function in $P(X,\phi)$ theories 
	are 
		\begin{equation}
		S^{(3)} = 
		\int \d^3 x \, \d\tau \; \Big\{ 
			a g_1 \zeta'^3 +a^2 g_2 \zeta \zeta'^2
			+ a^2 g_3 \zeta (\partial \zeta)^2
			+ a^2 g_4 \zeta' \partial_i \zeta
				\partial^i (\partial^{-2} \,\zeta')
			+ a^2 g_5 \partial^2 \zeta
				(\partial_i \partial^{-2}\,\zeta')^2	
		\Big\} \ .
		\label{eq:zeta-action3_chapter2}
	\end{equation}
The interaction coefficients $g_i$ can be read from
Eq. \eqref{eq:shiftaction}.

	The calculation we have described above follows 
the traditional methodology to compute the bispectrum in a given
inflationary theory: 
starting from the action for 
the background field $\phi$ we 
have applied perturbation theory to derive the action for
the 
perturbation $\zeta$.

However, recent developments have simplified this procedure. 
First, Gao \& Steer \cite{Gao:2011qe} and 
De Felice \& Tsujikawa \cite{DeFelice:2011uc} 
(see also Ref. \cite{RenauxPetel:2011sb})
obtained the universal action for perturbations in stable single-field models
involving what we call \textit{Horndeski operators} \cite{Deffayet:2011gz}
(cubic operators in $\zeta$). 
Over thirty years ago, Horndeski had already written the most general 
action involving one single scalar field 
yielding equations of motion which were at most 
second-order in derivatives \cite{Horndeski:1974}.
This action has the following structure
\begin{equation}
S=\int \d^4 x \sqrt{-g}\ 
\bigg\{
\dfrac{R}{2} +P_1(X,\phi) -P_2(X,\phi)\Box \phi 
+\mathcal{L}_3+\mathcal{L}_4
\bigg\}\ \ ,
\label{eq:horndeski-action}
\end{equation}
where 
\begin{subequations}
	\begin{align}
		\mathcal{L}_3 &= P_3(X,\phi)R+P_{3,X} \bigg[
		(\Box \phi)^2 -\big(\nabla_\mu \nabla_\nu \phi \big)\, 
		\big( \nabla^{\mu} \nabla^{\nu} \phi \big)
		\bigg] 
		\\
		\mathcal{L}_4 &= P_4(X,\phi) G_{\mu\nu} (\nabla^{\mu} \nabla^{\nu} \phi)
		-\dfrac{1}{6} P_{4,X}
		\bigg[
(\Box\phi)^3 -3 \Box \phi (\nabla_\mu \nabla_\nu \phi) (\nabla^{\mu} \nabla^{\nu} \phi)
\nonumber \\
& \hspace*{6cm}
+2 \, (\nabla^{\mu} \nabla_{\alpha}\phi) \, (\nabla^{\alpha} \nabla_{\beta}\phi)
(\nabla^{\beta}\nabla_{\mu} \phi)
		\bigg]\ \ , 
	\end{align}
	\label{eq:horndeski}
\end{subequations}
in which 
$P_1 , \cdots , P_4$ are arbitrary functions of $\phi$ and $X$, 
$P_{i,X}\equiv \partial P_i/\partial X$
and
$G_{\mu\nu}\equiv R_{\mu\nu}-\frac{1}{2}R g_{\mu\nu}$ 
is the Einstein tensor. 
Although the action involves high-order derivatives, 
the non-minimal couplings to the curvature ensure that the 
resulting Euler-Lagrange equations for $\phi$ are at most
second-order in derivatives. This is an essential requirement
for the theory to respect unitarity.
Recently there has been renewed interest in these theories
in the context of DGP
theories \cite{Dvali:2000hr}, 
and gravity theories based on galileon models (see, for example, 
Refs. \cite{Nicolis:2008in, Deffayet:2009wt, Deffayet:2009mn, deRham:2010eu};
see also Ref. \cite{Charmousis:2011bf}).

Unless an additional symmetry is 
imposed on these theories (for example, through the 
specific form of the functions $P_i(X,\phi)$), 
they will not be stable against radiative corrections.
Among higher-derivative theories, a notable exception
is DBI inflation \cite{Silverstein:2003hf}, 
where a higher dimensional boost protects the coefficients 
in the function $P(X,\phi)$ from receiving large radiative corrections.
If we, however, relax this requirement, 
we conclude that any healthy, single-field
inflation model can be written in the form \eqref{eq:horndeski-action}. 
The only model dependent features in the action would reside in different 
coefficients chosen in the functions $P_{i}(X,\phi)$.

Second, 
it was shown in Ref. \cite{RenauxPetel:2011sb} that the cubic action for $\zeta$ has a 
minimal representation in terms of only \textit{five} of these Horndeski operators.
This action was   
originally derived in Ref. \cite{Burrage:2011hd}, 
on which this chapter is based. It is our Eq. \eqref{eq:zeta-action3_chapter2}.

With these latest developments we arrive at a universal methodology 
to compute the bispectrum
of \emph{all} single-field models, 
with the cubic action for $\zeta$ being always of the form
	\begin{equation}
		S^{(3)} = 
		\int \d^3 x \, \d\tau \; \Big\{ 
			a \Lambda_1 \zeta'^3 +a^2 \Lambda_2 \zeta \zeta'^2
			+ a^2 \Lambda_3 \zeta (\partial \zeta)^2
			+ a^2 \Lambda_4 \zeta' \partial_i \zeta
				\partial^i (\partial^{-2}\zeta')
			+ a^2 \Lambda_5 \partial^2 \zeta
				(\partial_i \partial^{-2}\zeta')^2	
		\Big\} \ .
		\label{eq:zeta-action3-chapter2end}
	\end{equation}
The model-dependent imprints will be encoded in each of the 
\textit{five} coefficients $\Lambda_i$ of 
the Horndeski operators. There is a priori \textit{no} 
hierarchy between these coefficients, although specialization 
to different models can impose specific ratios between $\Lambda_i$ 
(as in DBI inflation).
The action above will be our starting point in computing the bispectrum
for all single-field models, whether or not the slow-roll 
approximation is invoked.
This is because the action \eqref{eq:zeta-action3-chapter2end}
is \textit{perturbative} in $\zeta$, 
but \textit{exact} in slow-roll.

\newpage
\thispagestyle{empty}
\newpage

\appendix

\thispagestyle{empty}

	\begin{titlepage}
	\vspace{3cm}
		\center

		\vspace*{6cm}
		
		{\Huge{ \bf Appendices}}
		
		\vspace{2cm}
\begin{quote}
\textit{Algebra is generous: she often gives more than is asked for.} 
\begin{flushright}
\textit{Jean d'Alembert}  (1717-1783)
\end{flushright}
\end{quote}
		
		\vspace{1,9cm}

		\end{titlepage}
		
	\newpage
\thispagestyle{empty}

\chapter{Secção de resultados}

\section{Olá}

	\bibliographystyle{JHEPmodplain}
	\bibliography{references}

\end{document}